\documentstyle[11pt]{article}

\font\fr=eufm10 scaled \magstep 1 
\font\es=msbm11                  
\newtheorem{teor}{Theorem}
\newtheorem{prop}{Proposition}

\newtheorem{lem}{Lemma}
\newtheorem{definition}{Definition}

\def\beq{\begin{equation}}
\def\eeq{\end{equation}}
\def\bea{\begin{eqnarray}}
\def\eea{\end{eqnarray}}
\def\beann{\begin{eqnarray*}}
\def\eeann{\end{eqnarray*}}
\def\beasn{\begin{sneqnarray}}
\def\eeasn{\end{sneqnarray}}
\def\ben{\begin{enumerate}}
\def\een{\end{enumerate}}
\def\bit{\begin{itemize}}
\def\eit{\end{itemize}}
\def\dst{\(\displaystyle}
\def\proof{ ({\sl Proof\/}) }
\def\derpar#1#2{\frac{\partial{#1}}{\partial{#2}}}

\def\feble#1{\mathrel{\mathop\simeq\limits_{#1}}}

\def\moment#1#2#3{{#1}_{#2}, \ldots, {#1}_{#3}}
\def\qed{\ifvmode\removelastskip\fi
{\unskip\nobreak\hfil\penalty50\hbox{}\nobreak\hfil
\hbox{\vrule height1.2ex width1.2ex}\parfillskip=0pt
\finalhyphendemerits=0 \par\smallskip}}

\def\vf{\mbox{\fr X}}
\def\df{{\mit\Omega}}
\def\Lag{{\cal L}}

\def\d{{\rm d}}

\def\Real{\mbox{\es R}}

\def\Tan{{\rm T}}

\def\inn{\mathop{i}\nolimits}
\def\Cinfty{{\rm C}^\infty}
\def\tabaddress#1{{\small\it\begin{tabular}[t]{c}#1
\\[1.2ex]\end{tabular}}}

\def\ls{(J^1E,\Omega_\Lag )}
\def\hs{(J^{1*}E,\Omega_h)}
\def\hso{({\cal P},\Omega_h^0)}

\title{LAGRANGIAN-HAMILTONIAN UNIFIED FORMALISM\\
  FOR FIELD THEORY.}
\author{\sc Arturo Echeverr\'{\i}a-Enr\'{\i}quez
  \\
  \tabaddress{Departamento de Matem\'atica Aplicada IV\\
  Edificio C-3, Campus Norte UPC\\
  C/ Jordi Girona 1. E-08034 Barcelona. Spain}
  \\
{\sc Carlos L\'opez \thanks{{\bf e}-{\it mail}:
  carlopez@posta.unizar.es}},
  \\
  \tabaddress{Departamento de Matem\'atica Aplicada, UZ\\
  C.P.S., C/  Mar\'{\i}a de Luna, 3. E-50015 Zaragoza. Spain}
  \\
{\sc Jes\'us Mar\'\i n-Solano \thanks{{\bf e}-{\it mail}:
  jmarin@ub.edu}},
  \\
  \tabaddress{Departamento de Matem\'atica Econ\'omica, Financiera
   y Actuarial, UB\\
   Av. Diagonal 690. E-08034 Barcelona. Spain}
    \\
{\sc Miguel C. Mu\~noz-Lecanda\thanks{{\bf e}-{\it mail}:
  matmcml@mat.upc.es}},
{\sc Narciso Rom\'an-Roy\thanks{{\bf e}-{\it mail}:
  matnrr@mat.upc.es}},
  \\
  \tabaddress{Departamento de Matem\'atica Aplicada IV\\
   Edificio C-3, Campus Norte UPC\\
   C/ Jordi Girona 1. E-08034 Barcelona. Spain}}
\date{{\sl J. Math. Phys.} {\bf 45}(1) (2004) 360-380.)}

\begin{document}

\maketitle

\pagestyle{myheadings}

\thispagestyle{empty}

\begin{abstract}
The {\sl Rusk-Skinner formalism} was developed in order to give a
geometrical unified formalism for describing mechanical systems.
It incorporates all the characteristics of Lagrangian and
Hamiltonian descriptions of these systems (including dynamical
equations and solutions, constraints, Legendre map, evolution
operators, equivalence, etc.).

In this work we extend this unified framework to first-order classical field theories,
and show how this description comprises the main features
of the Lagrangian and Hamiltonian formalisms,
both for the regular and singular cases. This formulation is a
first step toward further applications in optimal control theory
for PDE's.
\end{abstract}


\bigskip

\vbox{\raggedleft AMS s.\,c.\,(2000):
51P05, 53C05, 53C80, 55R10, 58A20, 58A30, 70S05. \\
PACS (1999): 02.40.Hw, 02.40.Vh, 11.10.Ef, 11.10 Kk. }\null

\markright{\sc A. Echeverr\'ia-Enr\'\i quez {\it et al\/},
    \sl Lagrangian-Hamiltonian unified formalism...}

  \clearpage

\section{Introduction}

  In ordinary autonomous classical theories in mechanics there is a unified formulation
of Lagrangian and Hamiltonian formalisms \cite{SR-83}, which is
based on the use of the {\sl Whitney sum} of the tangent and
cotangent bundles $W=\Tan Q\oplus\Tan^*Q\equiv\Tan
Q\times_Q\Tan^*Q$ (the {\sl velocity} and {\sl momentum phase
spaces} of the system). In this space, velocities and momenta are
independent coordinates. There is a canonical presymplectic form
$\Omega$ (the pull-back of the canonical form in $T^*Q$), and a
natural {\it coupling function}, locally expressed as  $p_iv^i$,
is defined by contraction between vectors and covectors. Given a
Lagrangian $L\in\Cinfty (\Tan Q)$, a Hamiltonian function, locally
given by  $H=p_iv^i -L(q,v)$, is determined, and, using the usual
constraint algorithm  for the geometric equation $i(X)\Omega = dH$
associated to the Hamiltonian system $(W, \Omega , H)$, we obtain
that:
 \ben
  \item The first constraint submanifold $W_1$ is
isomorphic to $TQ$, and the momenta ${{\partial L}\over{\partial
v^i}} = p_i$ are determined as constraints. \item The geometric
equation contains the second order condition $v^i = {{dq^i}\over
{dt}}$.
 \item The identification $W_1 \equiv TQ$ allows us to
recover the Lagrangian formalism.
 \item The projection to the
cotangent bundle generates the Hamiltonian formalism, including
constraints. The Legendre map and the time evolution operator are
straightforwardly obtained by the previous identification and
projection \cite{CL-87}.
 \een
  It is also worth noticing that this
space is also appropriate for the formulation of different kind of
problems in Optimal Control \cite{CM-2000}, \cite{CM-02}, \cite{CLMM-2002},
\cite{LCDM-02}, \cite{LM-2000}. Furthermore, in
\cite{CMC-2002} and \cite{LMM-2002} this unified formalism has
been extended for non-autonomous mechanical systems.

Our aim in this paper is to reproduce the same construction for
first-order field theories, generating a unified description of
Lagrangian and Hamiltonian formalisms and its correspondence,
starting from the multisymplectic description of such theories
(see, for instance, \cite{CCI-91}, \cite{EMR-96}, \cite{EMR-00},
\cite{GMS-97}, \cite{Go-91b}, \cite{HK-02}, \cite{HK-02b},
\cite{LMM-96}, \cite{Sd-95}, for some general references on this
formalism. See also \cite{Aw-92}, \cite{Gu-87}, \cite{Ka-98},
\cite{KT-79}, \cite{LMO-98}, \cite{LMS-2002}, \cite{No-93} for
other geometric formulations of field theories). As is shown
throughout the paper,
 characteristics analogous to those pointed out for mechanical
systems can be stated in this context. In \cite{LMM-2002}, a first
approach to this subject has been made, focusing mainly on the
constraint algorithm for the singular case.

The organization of the paper is as follows: Section 2 is devoted
to reviewing the main features of the multisymplectic description
of Lagrangian and Hamiltonian field theories. In Section 3 we
develop the unified formalism for field theories: starting from
the {\sl extended jet-multimomentum bundle} (the analogous to the
Whitney sum in mechanics), we introduce the so-called {\sl
extended Hamiltonian system} and state the field equations for
sections, $m$-vector fields, connections and jet fields in this
framework. It is also shown how the standard Lagrangian and
Hamiltonian descriptions are recovered from this unified picture.
As a typical example, the {\sl minimal surface problem} is
described in this formalism in Section 4. Finally, we include an
appendix where basic features about connections, jet fields and
$m$-vector fields are displayed.

  Throughout this paper
  $\pi\colon E\to M$ will be a fiber bundle ($\dim\, M=m$, $\dim\, E=N+m$),
  where $M$ is an oriented manifold with volume form $\omega\in\df^m(M)$.
  $\pi^1\colon J^1E\to E$ is the
  jet bundle of local sections of $\pi$, and
  $\bar\pi^1=\pi\circ\pi^1\colon J^1E \longrightarrow M$
  gives another fiber bundle structure.
  $(x^\alpha,y^A,v^A_\alpha)$ will denote natural local systems
  of coordinates in $J^1E$, adapted to the bundle $E\to M$
  ($\alpha = 1,\ldots,m$; $A= 1,\ldots,N$), and such that
  $\omega=\d x^1\wedge\ldots\wedge\d x^m\equiv\d^mx$.
  Manifolds are real, paracompact,
  connected and $C^\infty$. Maps are $C^\infty$. Sum over crossed repeated
  indices is understood.

\section{Geometric framework for classical field theories}

\subsection{Lagrangian formalism}
\protect\label{lagform}

(For details concerning the contents of this and the next section,
see for instance \cite{CCI-91}, \cite{EMR-96}, \cite{EMR-00}, 
\cite{GMS-97}, \cite{LMM-96}, \cite{Sd-95}, \cite{BSF-88}, 
\cite{EMR-98}, \cite{Gc-73}, \cite{PR-01}, \cite{Sa-89}, \cite{EMR-99b}. 
See also appendix \ref{mvfdm}).

A {\sl classical field theory} is described by giving a
{\sl configuration fiber bundle} $\pi\colon E\to M$
and a {\sl Lagrangian density}, which is
a $\bar\pi^1$-semibasic $m$-form on $J^1E$ usually written as
$\Lag =L \bar\pi^{1^*}\omega$, where $L\in\Cinfty (J^1E)$
is the {\sl Lagrangian function} determined by $\Lag$ and $\omega$.
The {\sl Poincar\'e-Cartan $m$ and $(m+1)$-forms}
associated with the Lagrangian density $\Lag$
are defined using the {\sl vertical endomorphism}
${\cal V}$ of the bundle $J^1E$ (see \cite{Sa-89})
$$
\Theta_{\Lag}:=\inn({\cal V})\Lag+\Lag\in\df^{m}(J^1E)
\quad ;\quad
\Omega_{\Lag}:= -\d\Theta_{\Lag}\in\df^{m+1}(J^1E)
$$
A {\sl Lagrangian system} is a couple $\ls$. It is {\sl regular} if
$\Omega_{\Lag}$ is a multisymplectic $(m+1)$-form
(a closed $m$-form, $m>1$, is called {\sl multisymplectic} if it is
$1$-nondegenerate; elsewhere it is {\sl pre-multisymplectic}).
In natural charts in $J^1E$ we have
\dst {\cal V}=(\d y^A-v^A_\alpha\d x^\alpha)\otimes\derpar{}{v^A_\nu}
\otimes\derpar{}{x^\nu}\), and
\beann
\Theta_{\Lag}&=&\derpar{L}{v^A_\mu}\d y^A\wedge\d^{m-1}x_\mu -
\left(\derpar{L}{v^A_\mu}v^A_\mu -L\right)\d^{m}x
\\
\Omega_{\Lag}&=&
-\frac{\partial^2L}{\partial v^B_\nu\partial v^A_\alpha}
\d v^B_\nu\wedge\d y^A\wedge\d^{m-1}x_\alpha
-\frac{\partial^2L}{\partial y^B\partial v^A_\alpha}\d y^B\wedge
\d y^A\wedge\d^{m-1}x_\alpha +
  \\  & &
+ \frac{\partial^2L}{\partial v^B_\nu\partial v^A_\alpha}v^A_\alpha
\d v^B_\nu\wedge\d^mx  +
\left(\frac{\partial^2L}{\partial y^B\partial v^A_\alpha}v^A_\alpha
  -\derpar{L}{y^B}+
\frac{\partial^2L}{\partial x^\alpha\partial v^B_\alpha}
\right)\d y^B\wedge\d^mx
\eeann
(where \dst\d^{m-1}x_\alpha\equiv\inn\left(\derpar{}{x^\alpha}\right)\d^mx\) );
  the regularity condition is equivalent to
\dst det\left(\frac{\partial^2L}
{\partial v^A_\alpha\partial v^B_\nu}(\bar y)\right)\not= 0\), for
every $\bar y\in J^1E$.

The {\sl Lagrangian problem} associated with a Lagrangian system
$\ls$ consists in finding sections $\phi\in\Gamma(M,E)$,
the set of sections of $\pi$, which are characterized by the condition
  $$
  (j^1\phi)^*\inn (X)\Omega_L=0 \quad ,\quad
  \mbox{\rm for every $X\in\vf (J^1E)$}
  $$
  In natural coordinates, if $\phi(x)=(x^\alpha,\phi^A(x))$,
  this condition is equivalent to demanding that
  $\phi$ satisfy the {\sl Euler-Lagrange equations}
  \beq
  \derpar{L}{y^A}\Big\vert_{j^1\phi} - \derpar{}{x^\alpha} \left(
\derpar{L}{v_\alpha^A}
  \right) \Big\vert_{j^1\phi} = 0
  \quad , \quad \mbox{\rm (for $A=1,\ldots ,N$)}
  \label{eleqs}
  \eeq

  The problem of finding these sections
  can be formulated equivalently as follows:
  finding a distribution $D$ of $\Tan (J^1E)$ such that
  it is integrable (that is, {\sl involutive\/}),
  $m$-dimensional, $\bar\pi^1$-transverse, and
  the integral manifolds of $D$ are the image of sections solution
  of the above equations (therefore, lifting of $\pi$-sections).
  This is equivalent to stating that the sections solution to
  the Lagrangian problem are the integral sections of one of the following
equivalent elements:
\bit
\item
A class of holonomic $m$-vector fields $\{ X_{\Lag}\}\subset\vf^m(J^1E)$, such that
 $\inn (X_{\Lag})\Omega_{\Lag}=0$, for every $X_\Lag\in\{ X_{\Lag}\}$.
\item
A holonomic connection $\nabla_\Lag$ in $\bar\pi^1\colon J¹E\to M$ such that
$\inn(\nabla_{\Lag})\Omega_{\Lag}=(m-1)\Omega_\Lag$.
\item
A holonomic jet field ${\mit\Psi}_\Lag\colon J^1E\to J^1J^1E$, such that
$\inn ({\mit\Psi}_{\Lag})\Omega_{\Lag}=0$
(the contraction of jet fields with differential forms
is defined in \cite{EMR-96}).
\eit
  Semi-holonomic locally decomposable $m$-vector fields,
  jet fields and connections which are solution to these equations
  are called {\sl Euler-Lagrange $m$-vector fields, jet fields}
  and {\sl connections} for $\ls$.
In a natural chart in $J^1E$, the local expressions of these elements are
\beann
  X_\Lag &=&
f \bigwedge_{\alpha=1}^m
\left(\derpar{}{x^\alpha}+F_\alpha^A\derpar{}{y^A}+
  G_{\alpha\nu}^A\derpar{}{v_\nu^A}\right)
  \\
\nabla_\Lag &=& \d x^\alpha\otimes
\left(\derpar{}{x^\alpha}+F_\alpha^A\derpar{}{y^A}+
  G_{\alpha\nu}^A\derpar{}{v_\nu^A}\right)
  \\
{\mit\Psi_\Lag}&=&
(x^\alpha,y^A,v^A_\alpha,F^A_\alpha,G^A_{\alpha\eta})
  \eeann
with $F^A_\alpha=v^A_\alpha$ (which is the local expression
of the semi-holonomy condition),
and where the coefficients
$G^A_{\alpha\nu}$ are related by the system of linear equations
  \beq
\frac{\partial^2L}{\partial v^A_\alpha\partial  v^B_\nu}G^A_{\alpha\nu}=
\derpar{L}{y^B}-\frac{\partial^2L}{\partial x^\nu\partial v^B_\nu}-
\frac{\partial^2L}{\partial y^A\partial v^B_\nu}v^A_\nu
\qquad (A,B=1,\ldots ,N)
  \label{eqsG0}
  \eeq
  $f\in\Cinfty(J^1E)$ is an arbitrary non-vanishing function.
  A representative of the class $\{ X_\Lag\}$ can be selected by the condition
  $\inn (X_\Lag)(\bar\pi^{1*}\omega)=1$, which leads to
  $f=1$ in the above local expression.
Therefore, if \dst j^1\phi=
\left( x^\mu,\phi^A,\derpar{\phi^A}{x^\nu}\right)\)
is an integral section of $X_\Lag$, then
\dst v^A_\alpha=\derpar{\phi^A}{x^\alpha}\) , and hence
the coefficients $G^B_{\alpha\nu}$ must satisfy the equations
$$
  G_{\nu\eta}^A\left(x^\alpha ,\phi^A,\derpar{\phi^A}{x^\alpha}\right) =
  \frac{\partial^2\phi^A}{\partial x^\eta\partial x^\nu}
\quad ;\quad
(A=1,\ldots ,N\ ;\ \eta,\nu=1,\ldots ,m)
$$
As a consequence, the system (\ref{eqsG0}) is equivalent
to the Euler-Lagrange equations (\ref{eleqs}) for $\phi$.

If $\ls$ is a regular Lagrangian system, the existence of classes
of Euler-Lagrange $m$-vector fields for $\Lag$ (or what is
equivalent, Euler-Lagrange jet fields or connections) is assured. For
singular Lagrangian systems, the existence of this kind of
solutions is not assured except perhaps on some submanifold
$S\hookrightarrow J^1E$. Furthermore, solutions of the field
equations can exist (in general, on some submanifold of $J^1E$),
but none of them are semi-holonomic (at any point of this
submanifold). In both cases, the integrability of these solutions
is not assured, except perhaps on a smaller submanifold $I$ such
that the integral sections are contained in $I$.

\subsection{Hamiltonian formalism}
\protect\label{hamform}

For the Hamiltonian formalism of field theories,
we have the {\sl extended multimomentum bundle} ${\cal M}\pi$,
which is the bundle of $m$-forms on
$E$ vanishing by contraction with two $\pi$-vertical vector fields
(or equivalently, the set of affine maps from
$J^1E$ to $\pi^*\Lambda^m\Tan^*M$ \cite{CCI-91}, \cite{GIMMSY-mm}),
and the {\sl restricted multimomentum bundle}
 $J^{1*}E\equiv{\cal M}\pi/\pi^*\Lambda^m\Tan^*M$.
 We have the natural projections
 $$
 \tau^1\colon J^{1*}E\to E \ ,\
 \bar\tau^1=\pi\circ\tau^1\colon J^{1*}E\to M \ ,\
 \mu \colon {\cal M}\pi\to J^{1*}E \ ,\
\hat\mu=\bar\tau^1\circ\mu\colon{\cal M}\pi\to M .
 $$
 Given a system of coordinates adapted to the bundle
 $\pi\colon E\to M$, we can construct natural coordinates
 $(x^\alpha,y^A,p^\alpha_A,p)$ ($\alpha = 1,\ldots,m$; $A= 1,\ldots,N$)
 in ${\cal M}\pi$, corresponding to the $m$-covector
 ${\bf p}=p\d^mx+p_A^{\alpha}\d y^A\wedge\d^{m-1}x_\alpha\in{\cal M}\pi$,
 and $(x^{\alpha},y^A,p_A^{\alpha})$ in $J^{1*}E$, for the class
 $[{\bf p}]=p_A^\alpha\d y^A\wedge\d^{m-1}x_\alpha+\langle\d^mx\rangle\in J^{1*}E$.

  Now, if $\ls$ is a Lagrangian system,
  the {\sl extended Legendre map} associated with $\Lag$,
  $\widetilde{{\cal F}\Lag}\colon J^1E\to {\cal M}\pi$, is defined as:
   \beq
  [\widetilde{{\cal F}\Lag}(\bar y)](\moment{Z}{1}{m}):=
  (\Theta_{\Lag})_{\bar y}(\moment{\bar Z}{1}{m})
  \label{elm}
  \eeq
  where $\moment{Z}{1}{m}\in\Tan_{\pi^1(\bar y)}E$, and
  $\moment{\bar Z}{1}{m}\in\Tan_{\bar y}J^1E$ are such that
  $\Tan_{\bar y}\pi^1\bar Z_\alpha=Z_\alpha$.
  Then the {\sl restricted Legendre map} associated with $\Lag$ is
  ${\cal F}\Lag :=\mu\circ\widetilde{{\cal F}\Lag}$.
  Their local expressions are
  $$
  \begin{array}{ccccccc}
  \widetilde{{\cal F}\Lag}^*x^\alpha = x^\alpha &\quad\ , \ \quad&
  \widetilde{{\cal F}\Lag}^*y^A = y^A &\quad\  , \quad&
  \widetilde{{\cal F}\Lag}^*p_A^\alpha =\derpar{L}{v^A_\alpha}
  &\quad\ , \quad&
  \widetilde{{\cal F}\Lag}^*p =L-v^A_\alpha\derpar{L}{v^A_\alpha}
  \\
  {\cal F}\Lag^*x^\alpha = x^\alpha &\quad\ , \ \quad&
  {\cal F}\Lag^*y^A = y^A &\quad\  , \quad&
  {\cal F}\Lag^*p_A^\alpha =\derpar{L}{v^A_\alpha} & &
  \end{array}
  $$
  Therefore, $\ls$ is a {\sl regular}
  Lagrangian system if ${\cal F}\Lag$ is a local diffeomorphism
  (this definition is equivalent to that given above).
  Elsewhere $\ls$ is a {\sl singular} Lagrangian system.
  As a particular case, $\ls$ is a {\sl hyper-regular}
  Lagrangian system if ${\cal F}\Lag$ is a global diffeomorphism.
  A singular Lagrangian system $\ls$ is {\sl almost-regular} if:
  ${\cal P}:={\cal F}\Lag (J^1E)$ is a closed submanifold of $J^{1*}E$
  (we will denote the natural imbedding by
   $\jmath_0\colon {\cal P}\hookrightarrow J^{1*}E$),
  ${\cal F}\Lag$ is a submersion onto its image, and
  for every $\bar y\in J^1E$, the fibres
  ${\cal F}\Lag^{-1}({\cal F}\Lag (\bar y))$
  are connected submanifolds of $J^1E$.

In order to construct a {\sl Hamiltonian system} associated with
$\ls$, recall that the multicotangent bundle $\Lambda^m\Tan^*E$ is
endowed with a natural canonical form ${\bf
\Theta}\in\df^m(\Lambda^m\Tan^*E)$, which is the tautological form
defined as follows: let $\tau_E\colon\Tan^*E\to E$ be the natural
projection, and $\Lambda^m\tau_E\colon\Lambda^m\Tan^*E\to E$ its
natural extension; then, for every ${\bf \bar
p}\in\Lambda^m\Tan^*E$ (where ${\bf \bar p}=(y,\beta)$, with $y\in
E$ and $\beta\in\Lambda^m\Tan_y^*E$), and for every
$\moment{X}{1}{m}\in\vf(\Lambda^m\Tan^*E)$ we have
$$
[{\bf \Theta}(\moment{X}{1}{m})]_{\bf \bar p}:=
[(\Lambda^m\tau_E)^*\beta](X_{1_{\bf \bar p}},\ldots ,X_{m_{\bf \bar p}})=
\beta(\Tan_{\bf \bar p}\Lambda^m\tau_E(X_{1_{\bf \bar p}}),\ldots ,
\Tan_{\bf \bar p}\Lambda^m\tau_E(X_{m_{\bf \bar p}}))
$$
Thus we also have the multisymplectic form ${\bf \Omega}:=-\d{\bf
\Theta}\in\df^{m+1}(\Lambda^m\Tan^*E)$.
  But ${\cal M}\pi\equiv\Lambda^m_1\Tan^*E$ is a subbundle of
  $\Lambda^m\Tan^*E$. Then, if
  $\lambda\colon\Lambda^m_1\Tan^*E\hookrightarrow\Lambda^m\Tan^*E$
  is the natural imbedding,
  $\Theta :=\lambda^* {\bf \Theta}$ and
  $\Omega :=-\d\Theta=\lambda^*{\bf \Omega}$
  are canonical forms in ${\cal M}\pi$, which are called
  the {\sl multimomentum Liouville} $m$ and $(m+1)$ {\sl forms}.
In particular, we have that
$\Theta({\bf p})=(\tau_1\circ\mu)^*{\bf p}$, for every ${\bf p}\in{\cal M}\pi$.
  Their local expressions are
  \beq
  \Theta = p_A^\alpha\d y^A\wedge\d^{m-1}x_\alpha+p\d^mx
  \quad , \quad
  \Omega = -\d p_A^\alpha\wedge\d y^A\wedge\d^{m-1}x_\alpha-\d p\wedge\d^mx
  \label{Omegah}
  \eeq
Observe that
  $\widetilde{{\cal F}\Lag}^*\Theta=\Theta_{\Lag}$,
  and $\widetilde{{\cal F}\Lag}^*\Omega=\Omega_{\Lag}$.

  Now, if $\ls$ is a hyper-regular Lagrangian system, then
  $\tilde{\cal P}:=\widetilde{{\cal F}\Lag}(J^1E)$ is a
  1-codimensional and $\mu$-transverse imbedded submanifold of ${\cal M}\pi$
  (we will denote the natural imbedding by
   $\tilde\jmath_0\colon\tilde{\cal P}\hookrightarrow{\cal M}\pi$),
  which is diffeomorphic to $J^{1*}E$. This diffeomorphism is $\mu^{-1}$, when $\mu$ is
  restricted to $\tilde{\cal P}$, and also coincides with the map
  $h:=\widetilde{{\cal F}\Lag}\circ{\cal F}\Lag^{-1}$,
  when it is restricted onto its image (which is just $\tilde{\cal P}$).
  This map $h$ is called a {\sl Hamiltonian section}, and
  can be used to construct the {\sl Hamilton-Cartan} $m$ and $(m+1)$ {\sl forms}
  of $J^{1*}E$ by making
  $$
  \Theta_h=h^*\Theta\in\df^m(J^{1*}E)
\quad , \quad
\Omega_h=h^*\Omega\in\df^{m+1}(J^{1*}E)
  $$
  The couple $\hs$ is said to be the {\sl Hamiltonian system}
  associated with the hyper-regular Lagrangian system $\ls$.
  Locally, the Hamiltonian section $h$ is specified by the
  {\sl local Hamiltonian function}
  $H=p^\alpha_A (F\Lag^{-1})^*v_\alpha^A-(F\Lag^{-1})^*L$,
  that is, $h(x^\alpha,y^A,p^\alpha_A)=(x^\alpha,y^A,p^\alpha_A,-H)$.
  Then we have the local expressions
  $$
  \Theta_h = p_A^\alpha\d y^A\wedge\d^{m-1}x_\alpha -H\d^mx
  \quad , \quad
  \Omega_h = -\d p_A^\alpha\wedge\d y^A\wedge\d^{m-1}x_\alpha +
  \d H\wedge\d^mx
  $$
   Of course
  ${\cal F}\Lag^*\Theta_h=\Theta_{\Lag}$,
  and ${\cal F}\Lag^*\Omega_h=\Omega_{\Lag}$.

  The {\sl Hamiltonian problem} associated with the Hamiltonian
  system $\hs$ consists in finding
  sections $\psi\in\Gamma(M,J^{1*}E)$,
  which are characterized by the condition
  $$
  \psi^*\inn (X)\Omega_h=0 \quad , \quad
  \mbox{\rm  for every $X\in\vf (J^{1*}E)$}
  $$
In natural coordinates, if
  $\psi(x)=(x^\alpha,y^A(x),p^\alpha_A(x))$, this condition leads to the
  so-called {\sl Hamilton-De Donder-Weyl equations} (for the section $\psi$).

 The problem of finding these sections
  can be formulated equivalently as follows:
  finding a distribution $D$ of $\Tan (J^{1*}E)$ such that
  $D$ is integrable (that is, {\sl involutive\/}),
  $m$-dimensional, $\bar\tau^1$-transverse, and
  its integral manifolds are the sections solution to the
  above equations. This is equivalent to stating that the sections solution to
  the Hamiltonian problem are the integral sections of one of the following
  equivalent elements:
\bit
\item
A class of integrable and $\bar\tau^1$-transverse $m$-vector fields
  $\{ X_{\cal H}\}\subset\vf^m(J^{1*}E)$ satisfying that
  $\inn (X_{\cal H})\Omega_h=0$, for every $X_{\cal H}\in\{ X_{\cal H}\}$.
\item
An integrable connection $\nabla_{\cal H}$
in $\bar\tau^1\colon J^{1*}E\to M$ such that
$\inn(\nabla_{\cal H})\Omega_h=(m-1)\Omega_h$.
\item
An integrable jet field
  ${\mit\Psi}_{\cal H}\colon J^{1*}E\to J^1J^{1*}E$, such that
$\inn ({\mit\Psi}_{\cal H})\Omega_h=0$.
\eit
$\bar\tau^1$-transverse and locally decomposable $m$-vector fields,
  orientable jet fields and orientable connections
  which are solutions of these equations
  are called {\sl Hamilton-De Donder-Weyl (HDW) $m$-vector fields,
  jet fields} and {\sl connections} for $\hs$.
Their local expressions in natural coordinates are
\beann
   X_{{\cal H}} &=& f \bigwedge_{\alpha=1}^m
  \left(\derpar{}{x^\alpha}+F_\alpha^A\derpar{}{y^A}+
  G^\eta_{A\alpha}\derpar{}{p^\eta_A}\right)
   \\
{\mit\Psi}_{{\cal H}} &=&
(x^\alpha,y^A,p_A^\alpha,;F^A_\alpha,G^\eta_{A\alpha})
  \\
\nabla_{{\cal H}} &=& \d x^\alpha\otimes
\left(\derpar{}{x^\alpha}+F_\alpha^A\derpar{}{y^A}+
  G_{A\alpha}^\nu\derpar{}{p^\nu_A}\right)
  \eeann
  where $f\in\Cinfty(J^{1*}E)$ is a non-vanishing function,
  and the coefficients $F_\alpha^A,G^\eta_{A\alpha}$ are related by the
  system of linear equations
  $$
  F^A_\alpha=\derpar{H}{p_A^\alpha}
  \quad , \quad
  G^\nu_{A\nu}=-\derpar{H}{y^A}
  $$
  Now, if $\psi(x)=(x^\alpha ,y^A(x)=\psi^A(x),p^\alpha_A(x)=\psi^\alpha_A(x))$
  is an integral section of $X_{\cal H}$ then
  $$
  \derpar{H}{p^\alpha_A}\Bigg\vert_{\psi}=
  F^A_\alpha\circ \psi =\derpar{\psi^A}{x_\alpha} \quad ; \quad
  -\derpar{H}{y^A}\Bigg\vert_{\psi}=
  G^\alpha_{A\alpha}\circ\psi = \derpar{\psi^\alpha_A}{x^\alpha}
  $$
  which are the Hamilton-De Donder-Weyl equations for $\psi$.
  As above, a representative of the class $\{ X_{\cal H}\}$
  can be selected by the condition
  $\inn (X_{\cal H})(\bar\tau^{1*}\omega)=1$, which leads to $f=1$
  in the above local expression.
  The existence of classes of HDW $m$-vector fields,
  jet fields and connections is assured.

  In an analogous way, if $\ls$ is an almost-regular Lagrangian system,
 the submanifold $\jmath\colon {\cal P}\hookrightarrow J^{1*}E$,
 is a fibre bundle over $E$ and $M$.
 In this case the $\mu$-transverse submanifold
 $\tilde{\cal P}\hookrightarrow{\cal M}\pi$ is diffeomorphic to
 ${\cal P}$. This diffeomorphism is denoted by
 $\tilde\mu\colon\tilde{\cal P}\to{\cal P}$,
 and it is just the restriction of the projection $\mu$ to $\tilde{\cal P}$.
 Then, taking the Hamiltonian section $\tilde h:=\tilde\jmath\circ\tilde\mu^{-1}$,
 we define the Hamilton-Cartan forms
  $$
 \Theta^0_h=\tilde h^*\Theta
 \quad ; \quad
 \Omega^0_h=\tilde h^*\Omega
 $$
 which verify that
 ${\cal F}\Lag_0^*\Theta^0_h=\Theta_{\Lag}$ and
 ${\cal F}\Lag_0^*\Omega^0_h=\Omega_{\Lag}$
(where ${\cal F}\Lag_0$ is the restriction map of
${\cal F}\Lag$ onto ${\cal P}$).
 Then $\hso$ is the {\sl Hamiltonian system}
 associated with the almost-regular Lagrangian system $\ls$,
 and we have the following diagram
 \beq
\begin{array}{cccc}
\begin{picture}(15,52)(0,0)
\put(0,0){\mbox{$J^1E$}}
\end{picture}
&
\begin{picture}(65,52)(0,0)
 \put(15,28){\mbox{$\widetilde{{\cal F}\Lag}_0$}}
 \put(24,-9){\mbox{${\cal F}\Lag_0$}}
 \put(0,7){\vector(2,1){65}}
 \put(0,2){\vector(1,0){65}}
\end{picture}
&
\begin{picture}(90,52)(0,0)
 \put(5,0){\mbox{${\cal P}$}}
 \put(5,42){\mbox{$\tilde{\cal P}$}}
 \put(5,13){\vector(0,1){25}}
 \put(10,38){\vector(0,-1){25}}
 \put(-15,20){\mbox{$\tilde\mu^{-1}$}}
 \put(12,22){\mbox{$\tilde\mu$}}
 \put(30,45){\vector(1,0){55}}
 \put(30,2){\vector(1,0){55}}
 \put(30,8){\vector(2,1){55}}
 \put(53,-9){\mbox{$\jmath$}}
 \put(48,33){\mbox{$\tilde\jmath$}}
 \put(65,12){\mbox{$\tilde h$}}
 \end{picture}
&
\begin{picture}(15,52)(0,0)
 \put(0,0){\mbox{$J^{1*}E$}}
 \put(0,41){\mbox{${\cal M}\pi$}}
 \put(10,38){\vector(0,-1){25}}
 \put(0,22){\mbox{$\mu$}}
\end{picture}
\\
& &
\begin{picture}(90,35)(0,0)
 \put(10,35){\vector(1,-1){35}}
 \put(5,11){\mbox{$\bar\tau^1_0$}}
 \put(90,11){\mbox{$\bar\tau^1$}}
 \put(100,35){\vector(-1,-1){35}}
\end{picture}
 &
\\
& & \quad\quad M &
 \end{array}
\label{arhs}
 \eeq
  Then, the {\sl Hamiltonian problem} associated with the Hamiltonian
  system $({\cal P},\Omega_h^0)$, and the equations for the sections
  of $\Gamma(M,{\cal P})$ solution to
  the Hamiltonian problem are stated as in the regular case.
  Now, the existence of the corresponding Hamilton-De Donder-Weyl $m$-vector fields,
  jet fields and connections for $({\cal P},\Omega_h^0)$ is not assured,
  except perhaps on some submanifold
  $P$ of ${\cal P}$, where the solution is not unique.

 From now on we will consider only regular or almost-regular systems.

\section{Unified formalism}

\subsection{Extended Hamiltonian system}

Given a fiber bundle $\pi:E \to M$ over an oriented manifold $(M,\omega )$,
we define the {\sl extended jet-multimomentum bundle} ${\cal W}$ and
the {\sl restricted jet-multimomentum bundle} ${\cal W}_r$ as
$$
{\cal W}:= J^1E \times_E {\cal M}\pi \quad , \quad
{\cal W}_r := J^1E\times_E J^{1*}E
$$
whose natural coordinates are $(x^{\alpha},y^A,v^A_{\alpha},p_A^{\alpha},p)$
and $(x^{\alpha},y^A,v^A_{\alpha},p_A^{\alpha})$, respectively.
We have the natural projections (submersions)
\bea
\rho_1\colon{\cal W}\to J^1E \ ,\
\rho_2\colon{\cal W}\to {\cal M}\pi \ ,\
\rho_E\colon{\cal W}\to E \ ,\
\rho_M\colon{\cal W}\to M
\label{project}
\\
\rho_1^r\colon{\cal W}_r\to J^1E \ ,\
\rho_2^r\colon{\cal W}_r\to J^{1*}E \ ,\
\rho_E^r\colon{\cal W}_r\to E \ ,\
\rho_M^r\colon{\cal W}_r\to M
\nonumber
\eea
  Note that $\pi^1\circ\rho_1=\tau^1\circ\mu\circ\rho_2=\rho_E$.
In addition, there is also the natural projection
$$
\begin{array}{ccccc}
\mu_{\cal W}&\colon& {\cal W}& \to &{\cal W}_r\\
& & (\bar y,{\bf p}) & \mapsto & (\bar y,[{\bf p}])
\end{array}
$$
The bundle ${\cal W}$ is endowed with the following canonical structures:

\begin{definition}
\ben
\item
The {\rm coupling $m$-form} in ${\cal W}$, de\-no\-ted by
${\cal C}$, is an $m$-form along $\rho_M$
which is defined as follows:
for every $\bar y\in J_y^1E$, with $\bar\pi^1(\bar y)=\pi(y)=x\in E$,
and ${\bf p}\in{\cal M}_y\pi$, let $w\equiv (\bar y,{\bf p})\in{\cal W}_y$, then
$$
{\cal C}(w):=(\Tan_x\phi)^*{\bf p}
$$
where $\phi\colon M\to E$ satisfies that $j^1\phi (x) = \bar y$.

Then, we denote by $\hat{\cal C}\in\df^m({\cal W})$ the
$\rho_M$-semibasic form associated with ${\cal C}$.
\item
The {\rm canonical $m$-form}
$\Theta_{\cal W}\in\df^m({\cal W})$ is defined by
$\Theta_{\cal W}:=\rho_2^*\Theta$, and it is therefore $\rho _E$-semibasic.

The {\rm canonical $(m+1)$-form} is the pre-multisymplectic form
$\Omega_{\cal W}:=-\d\Theta_{\cal W}=\rho_1^*\Omega\in\df^{m+1}({\cal W})$.
\een
\label{coupling}
\end{definition}

Being $\hat{\cal C}$  a $\rho_M$-semibasic form,
  there is $\hat C\in\Cinfty ({\cal W})$ such that
$\hat{\cal C}=\hat C(\rho_M^*\omega)$. Note also that  $\Omega_{\cal W}$
is not  1-nondegenerate, its kernel being the
$\rho _2$-vertical vectors; then, we call $({\cal W},
\Omega_{\cal W} )$ a pre-multisymplectic structure. This definition of the
coupling form is in fact an alternative
(obviously equivalent) presentation of the extended multimomentum
bundle as the set of affine maps from the jet
bundle $J^1E$ to $\pi$-basic $m$-forms.

The local expressions for $\Theta_{\cal W}$ and $\Omega_{\cal W}$ are the same as
(\ref{Omegah}), and for $\hat{\cal C}$ we have
$$
\hat{\cal C}(w)=(p+p_A^\alpha v^A_\alpha)\d^mx
$$

Given a Lagrangian density $\Lag\in\df^m(J^1E)$,
we denote $\hat\Lag:=\rho_1^*\Lag\in\df^m({\cal W})$,
and we can write $\hat\Lag=\hat L(\rho_M^*\omega)$,
with $\hat L=\rho_1^*L\in\Cinfty ({\cal W})$.
We define a {\sl Hamiltonian submanifold}
$$
{\cal W}_0:=\{ w\in{\cal W}\ | \ \hat\Lag(w)=\hat{\cal C}(w) \}
$$
So, ${\cal W}_0$ is the submanifold of $\cal W$ defined by the
constraint function
$\hat C-\hat L=0$. In local coordinates this constraint function is
$$
p + p_A^\alpha v^A_\alpha-\hat L(x^\nu,y^B,v^B_\nu)=0
$$
We have the natural imbedding
$\jmath_0\colon{\cal W}_0\hookrightarrow{\cal W}$,
as well as the projections (submersions)
$$
\rho_1^0\colon{\cal W}_0\to J^1E \ ,\
\rho_2^0\colon{\cal W}_0\to {\cal M}\pi \ ,\
\rho_E^0\colon{\cal W}_0\to E \ ,\
\rho_M^0\colon{\cal W}_0\to M
$$
which are the restrictions to ${\cal W}_0$ of the projections
(\ref{project}), and $\hat {\rho }_2^0 = \mu
\circ \rho _2^0 : {\cal W}_0 \to J^{1*}E$.
So we have the following diagram
$$
\begin{array}{ccc}
& \begin{picture}(135,20)(0,0)
\put(58,5){\mbox{${J^1E}$}}
\end{picture} &
\\
& \begin{picture}(135,30)(0,0)
\put(12,16){\mbox{$\rho_1^0$}}
\put(-3,-10){\vector(3,2){60}}
\put(55,10){\mbox{$\rho_1$}}
\put(69,-11){\vector(0,1){45}}
\put(141,-10){\vector(-3,2){60}}
\put(118,16){\mbox{$\rho_1^r$}}
\end{picture}
\\
{\cal W}_0 &
\begin{picture}(135,20)(0,0)
\put(28,10){\mbox{$\jmath_0$}}
\put(0,3){\vector(1,0){58}}
\put(64,0){\mbox{${\cal W}$}}
\put(100,10){\mbox{$\mu_{\cal W}$}}
\put(80,3){\vector(1,0){58}}
\end{picture}
&  {\cal W}_r
\\ &
\begin{picture}(135,100)(0,0)
\put(31,84){\mbox{$\rho_2^0$}}
\put(55,84){\mbox{$\rho_2$}}
\put(94,82){\mbox{$\rho_2^r$}}
\put(4,55){\mbox{$\hat\rho_2^0$}}
 \put(123,55){\mbox{$\hat\rho_2^r$}}
 \put(60,30){\mbox{$\mu$}}
\put(60,55){\mbox{${\cal M}\pi$}}
 \put(55,0){\mbox{$J^{1*}E$}}
\put(70,100){\vector(0,-1){35}}
\put(0,100){\vector(3,-2){55}}
 \put(140,100){\vector(-3,-2){55}}
\put(-5,100){\vector(2,-3){55}}
 \put(144,100){\vector(-2,-3){55}}
\put(70,48){\vector(0,-1){35}}
\end{picture} &
\end{array}
$$
Local coordinates in ${\cal W}_0$ are
$(x^{\alpha},y^A,v^A_\alpha,p_A^\alpha)$, and we have that
\beann
\rho_1^0(x^{\alpha},y^A,v^A_\alpha,p_A^\alpha) &=& (x^{\alpha},y^A,v^A_\alpha) \\
\jmath_0(x^{\alpha},y^A,v^A_\alpha,p_A^\alpha) &=&
   (x^{\alpha},y^A,v^A_\alpha,p_A^\alpha,L-v^A_\alpha p_A^\alpha) \\
\rho_2^0(x^{\alpha},y^A,v^A_\alpha,p_A^\alpha) &=&
(x^{\alpha},y^A,p_A^\alpha,L-v^A_\alpha p_A^\alpha) \\
\hat\rho_2^0(x^{\alpha},y^A,v^A_\alpha,p_A^\alpha) &=& (x^{\alpha},y^A,p_A^\alpha)
\eeann

\begin{prop}
${\cal W}_0$ is a $1$-codimensional $\mu_{\cal W}$-transversal submanifold
of ${\cal W}$, diffeomorphic to ${\cal W}_r$.
\label{1}
\end{prop}
\proof
For every $(\bar y,{\bf p})\in{\cal W}_0$, we have
$L(\bar y)\equiv\hat L(\bar y,{\bf p})=\hat C(\bar y,{\bf p})$, and
$$
(\mu_{\cal W}\circ\jmath_0)(\bar y,{\bf p})=\mu_{\cal W}(\bar y,{\bf p})=
(\bar y,\mu({\bf p}))=(\bar y,[{\bf p}])
$$

First, $\mu_{\cal W}\circ\jmath_0$ is injective:
let $(\bar y_1,{\bf p}_1),(\bar y_2,{\bf p}_2)\in{\cal W}_0$, then we have
$$
(\mu_{\cal W}\circ\jmath_0)(\bar y_1,{\bf p}_1)=(\mu_{\cal W}\circ\jmath_0)(\bar y_2,{\bf p}_2)
\,\Rightarrow\, (\bar y_1,\mu({\bf p}_1))=(\bar y_2,\mu({\bf p}_2))
\, \Rightarrow\,
\bar y_1=\bar y_2\ ,\ \mu({\bf p}_1)=\mu({\bf p}_2)
$$
hence
$$
L(\bar y_1)=L(\bar y_2)=\hat C(\bar y_1,{\bf p}_1)=
\hat C(\bar y_2,{\bf p}_2)
$$
In a local chart, third equality gives
$$
p({\bf p}_1)+p_A^\alpha({\bf p}_1)v^A_\alpha(\bar y_1)=
p({\bf p}_2)+p_A^\alpha({\bf p}_2)v^A_\alpha(\bar y_2)
$$
but $\mu({\bf p}_1)=\mu({\bf p}_2)$ implies that
$$
p_A^\alpha({\bf p}_1)=p_A^\alpha([{\bf p}_1])=p_A^\alpha([{\bf p}_2])=p_A^\alpha({\bf p}_2)
$$
therefore $p({\bf p}_1)=p({\bf p}_2)$ and hence ${\bf p}_1={\bf p}_2$.

Second, $\mu_{\cal W}\circ\jmath_0$ is onto:
Let $(\bar y,{\bf p})\in{\cal W}_r$, then there exists
$(\bar y,{\bf q})\in\jmath_0({\cal W}_0)$ such that
$[{\bf q}]=[{\bf p}]$. In fact, it suffices to take $[{\bf q}]$ in such a way that,
in a local chart of $J^1E\times_E{\cal M}\pi={\cal W}$
$$
p_A^\alpha({\bf q})=p_A^\alpha([{\bf p}]) \ , \
p({\bf q})=p_A^\alpha([{\bf p}])v^A_\alpha(\bar y)-L(\bar y)
$$

Finally, observe that ${\cal W}_0$ is defined  by the constraint
function $\hat L-\hat C$ and, as \dst\ker\,\mu_{\cal
W*}=\left\{\derpar{}{p}\right\}\) and  \dst\derpar{}{p}(\hat
L-\hat C)=1\), then ${\cal W}_0$ is a $1$-codimensional
submanifold of ${\cal W}$
 and $\mu_{\cal W}$-transversal.
\qed

As a consequence of this property, the
submanifold ${\cal W}_0$ induces a section
$\hat h\colon{\cal W}_r\to{\cal W}$ of the projection $\mu_{\cal W}$.
Locally, $\hat h$ is specified by
giving the local {\sl Hamiltonian function}
$\hat H= -\hat L+p_A^\alpha v^A_\alpha$; that is,
$\hat h(x^\alpha,y^A,v^A_\alpha,p^\alpha_A)=
(x^\alpha,y^A,v^A_\alpha,p^\alpha_A,-\hat H)$.
In this sense, $\hat h$ is said to be a {\sl Hamiltonian section} of
$\mu_{\cal W}$.

\textbf{Remark}:
It is important to point out that, from every
Hamiltonian $\mu_{\cal W}$-section
 $\hat h\colon{\cal W}_r\to{\cal W}$ in the extended unified formalism,
 we can recover a Hamiltonian $\mu$-section
 $\tilde h\colon{\cal P}\to{\cal M}\pi$ in the standard Hamiltonian formalism.
In fact, given $[{\bf p}]\in J^{1*}E$, the section $\hat h$ maps
every point
 $(\bar y,[{\bf p}])\in(\rho_2^r)^{-1}([{\bf p}])$ into
 $\rho_2^{-1}[\rho_2(\hat h(\bar y,[{\bf p}]))]$.
 So, the crucial point is the projectability of the local function
 $\hat H$ by $\rho_2$. But, being
 \dst\derpar{}{v^A_\alpha}\) a local basis for $\ker\,\rho_{2*}$,
 $\hat H$ is $\rho_2$-projectable iff
  \dst p_A^\alpha=\derpar{L}{v_A^\alpha}\), and this condition
  is fulfilled when
  $[{\bf p}]\in{\cal P}={\rm Im}\,{\cal F}\Lag\subset J^{1*}E$,
  which implies that
  $\rho_2[\hat h(\rho_2^r)^{-1})([{\bf p}]))]\in
  \tilde{\cal P}={\rm Im}\, \widetilde{{\cal F}\Lag}\subset{\cal M}\pi$.
  Hence, the Hamiltonian section $\tilde h$ is defined as follows
  $$
\tilde h([{\bf p}])=(\rho_2\circ\hat h)[(\rho_2^r)^{-1}(\jmath([{\bf p}]))]
\ ,\
\mbox{\rm for every $[{\bf p}]\in{\cal P}$}
 $$
So we have the diagram (see also diagram (\ref{arhs}))
 $$
\begin{picture}(90,52)(0,0)
 \put(5,0){\mbox{${\cal P}$}}
 \put(5,42){\mbox{$\tilde{\cal P}$}}
 \put(5,13){\vector(0,1){25}}
 \put(10,38){\vector(0,-1){25}}
 \put(-15,20){\mbox{$\tilde\mu^{-1}$}}
 \put(15,22){\mbox{$\tilde\mu$}}
 \put(20,45){\vector(1,0){55}}
 \put(20,2){\vector(1,0){55}}
 \put(20,8){\vector(2,1){55}}
 \put(53,-9){\mbox{$\jmath$}}
 \put(48,33){\mbox{$\tilde\jmath$}}
 \put(62,15){\mbox{$\tilde h$}}
 \end{picture}
\begin{picture}(90,52)(0,0)
 \put(-3,0){\mbox{$J^{1*}E$}}
 \put(0,42){\mbox{${\cal M}\pi$}}
 \put(10,38){\vector(0,-1){25}}
 \put(15,22){\mbox{$\mu$}}
 \put(85,45){\vector(-1,0){55}}
 \put(85,2){\vector(-1,0){55}}
 \put(53,-9){\mbox{$\rho_2^r$}}
 \put(48,33){\mbox{$\rho_2$}}
 \end{picture}
\begin{picture}(15,52)(0,0)
 \put(0,0){\mbox{${\cal W}_r$}}
 \put(0,41){\mbox{${\cal W}$}}
 \put(5,13){\vector(0,1){25}}
 \put(10,22){\mbox{$\hat h$}}
\end{picture}
 $$
(For (hyper) regular systems this diagram is the same with ${\rm
Im}\,{\cal F}\Lag=J^{1*}E$).

 Finally, we can define the forms
  $$
\Theta_0:=\jmath_0^*\Theta_{\cal W} = \rho _2^{0*} \Theta \in\df^m({\cal W}_0)
\quad , \quad
\Omega_0:=\jmath_0^*\Omega_{\cal W} =\rho _2^{0*} \Omega
\in\df^{m+1}({\cal W}_0)
$$
with local expressions
\bea
\Theta_0&=&(L-p_A^\alpha v^A_\alpha)\d^mx+
p_A^\alpha\d y^A\wedge\d^{m-1}x_\alpha
\nonumber \\
\Omega_0&=&\d (p_A^\alpha v^A_\alpha-L)\wedge\d^mx-
\d p_A^\alpha\wedge\d y^A\wedge\d^{m-1}x_\alpha
\label{coor0}
\eea
and we have obtained a (pre-multisymplectic) Hamiltonian system
$({\cal W}_0,\Omega_0)$, or equivalently $({\cal W}_r,\hat h^*\Omega_0)$.

\subsection{The field equations for sections}

The {\sl Lagrange-Hamiltonian problem} associated with
the system $({\cal W}_0,\Omega_0)$ consists in finding sections
$\psi_0\in\Gamma(M,{\cal W}_0)$
which are characterized by the condition
  \beq
  \psi_0^*\inn(Y_0)\Omega_0=0 \quad ,\quad
  \mbox{\rm for every $Y_0\in\vf({\cal W}_0)$}
  \label{psi0}
\eeq
 This equation gives different kinds of information, depending
on the type of the vector fields $Y_0$ involved. In particular,
using vector fields $Y_0$ which are $\hat\rho_2^0$-vertical, we
have:

\begin{lem}
If $Y_0\in\vf^{{\rm V}(\hat\rho_2^0)}({\cal W}_0)$
(i.e., $Y_0$ is $\hat\rho_2^0$-vertical),
 then $\inn(Y_0)\Omega_0$ is $\rho_M^0$-semibasic.
\end{lem}
\proof
A simple calculation in coordinates leads
to this result. In fact,
taking \dst\left\{\derpar{}{v^A_\alpha}\right\}\)
as a local basis for the $\hat\rho_2^0$-vertical vector fields,
and bearing in mind (\ref{coor0}) we obtain
$$
\inn\left(\derpar{}{v^A_\alpha}\right)\Omega_0=
\left( p_A^\alpha-\derpar{L}{v^A_\alpha}\right)\d^mx
$$
which are obviously $\rho_M^0$-semibasic forms.
\qed

As an immediate consequence,
 when $Y_0\in\vf^{{\rm V}(\hat\rho_2^0)}({\cal W}_0)$,
condition (\ref{psi0}) does not depend on the derivatives of
$\psi_0$: is a pointwise (algebraic) condition. We can define the
submanifold
$$
{\cal W}_1 = \{ (\bar y,{\bf p})\in{\cal W}_0 \ | \
\inn(V_0)(\Omega_0)_{({\bar y},{\bf p})}=0,\
\mbox{for every $V_0\in{\rm V}(\hat\rho_2^0 )$} \}
$$
which is called the {\sl first constraint submanifold}
of the Hamiltonian pre-multisymplectic system $({\cal W}_0,\Omega _0)$, as
every section $\psi_0$ solution to (\ref{psi0}) must take values in  ${\cal W}_1$.
We denote by $\jmath_1\colon{\cal W}_1\hookrightarrow{\cal W}_0$
the natural embedding.

Locally, ${\cal W}_1$ is defined in ${\cal W}_0$ by the constraints
\dst p^\alpha_A=\derpar{L}{v^A_\alpha}\). Moreover:

\begin{prop}
${\cal W}_1$ is the graph of $\widetilde{\cal FL}$; that is,
${\cal W}_1=\{ (\bar {y},\widetilde{\cal FL}(\bar y))\in{\cal W}\  \mid \
\bar y\in J^1E\}$.
\end{prop}
\proof
 Consider $\bar y\in J^1E$, let $\phi\colon M\to E$ be a
representative of $\bar y$, and ${\bf p}=\widetilde{\cal FL}(\bar
y)$. For every $U\in\Tan_{\bar\pi^1(\bar y)}M$, consider
$V=\Tan_{\bar\pi^1(\bar y)}\phi(U)$ and its canonical lifting
$\bar V=\Tan_{\bar\pi^1(\bar y)}j^1\phi(U)$. From the definition
of the extended Legendre map (\ref{elm}) we have that $(\Tan_{\bar
y}\pi)^*(\widetilde{{\cal F}\Lag}(\bar y))=(\Theta_\Lag)_{\bar
y}$, then
$$
\inn(\bar V)[(\Tan_{\bar y}\pi^1)^*(\widetilde{{\cal F}\Lag}(\bar y))]=
\inn(\bar V)(\Theta_\Lag)_{\bar y}
$$
Furthermore, as ${\bf p}=\widetilde{\cal FL}(\bar y)$, we also have that
\beann
\inn(\bar V)[(\Tan_{\bar y}\pi^1)^*(\widetilde{{\cal F}\Lag}(\bar y))]&=&
\inn(\Tan_{\bar\pi^1(\bar y)}j^1\phi(U))[(\Tan_{\bar y}\pi^1)^*{\bf p})=
\inn(\Tan_{\pi^1(\bar y)}[(\Tan_{\bar\pi^1({\bar y})}j^1\phi(U)]){\bf p}
\nonumber \\ &=&
\inn(\Tan_{\bar\pi^1(\bar y)}\phi(U)){\bf p}=
\inn(V){\bf p}
\eeann
Therefore we obtain
$$
\inn(U)(\phi^*{\bf p})=\inn(U)[(j^1\phi)^*(\Theta_\Lag)_{\bar y}]
$$
and bearing in mind the definition of the coupling form $\cal C$,
this condition becomes
$$
\inn(U)({\cal C}(\bar y,{\bf p}))=\inn(U)[(j^1\phi)^*\Theta_{\cal L})_{\bar y}]
$$
Since it holds for every $U\in\Tan_{\bar\pi^1(\bar y)}M$,
we conclude that
${\cal C}(\bar y,{\bf p})=[(j^1\phi)^*\Theta_{\cal L}]_{\bar y}$,
or equivalently, $\hat{\cal C}(\bar y,{\bf p})=\hat L({\bar y,{\bf p})}$,
where we have made use of the fact that $\Theta_{\cal L}$ is the sum
of the Lagrangian density $\Lag$ and a contact form
$\inn({\cal V}){\cal L}$ (vanishing by pull-back of lifted sections).
This is the condition defining ${\cal W}_0$,
and thus we have proved that $(\bar y,\widetilde{\cal FL}(\bar y))\in{\cal W}_0$,
for every $\bar y\in J^1E$; that is,
${\rm graph}\,\widetilde{{\cal F}\Lag}\subset{\cal W}_0$.
Furthermore, ${\rm graph}\,\widetilde{{\cal F}\Lag}$ and ${\cal W}_1$
are defined as subsets of ${\cal W}_0$ by the same local conditions:
\dst p_A^\alpha -\derpar{L}{v^A_\alpha} =0\).
So we conclude that
${\rm graph}\,\widetilde{{\cal F}\Lag}={\cal W}_1$.
\qed

Being ${\cal W}_1$ the graph of $\widetilde{\cal FL}$, it is diffeomorphic to
$J^1E$. Every section $\psi_0\colon M \to {\cal W}_0$ is of the form
$\psi_0=(\psi_\Lag,\psi_{\cal H})$, with $\psi_\Lag=\rho_1^0\circ\psi_0\colon M \to J^1E$,
and if $\psi_0$ takes values in ${\cal W}_1$ then
$\psi_{\cal H} =\widetilde{\cal FL}\circ\psi_\Lag$. In this way,
every constraint, differential equation, etc.
in the unified formalism can be translated to the
Lagrangian or the Hamiltonian formalisms by restriction to the first
or the second factors of the product bundle.

However, as was pointed out before, the geometric condition
(\ref{psi0}) in ${\cal W}_0$, which can be solved only for
sections $\psi_0\colon M\to{\cal W}_1\subset{\cal W}_0$, is
stronger than the Lagrangian condition
$\psi_\Lag^*\inn(Z)\Omega_\Lag= 0$, (for every $Z\in\vf(J^1E)$) in
$J^1E$, which can be translated to ${\cal W}_1$ by the natural
diffeomorphism between them. The reason is that $\Tan_{{\cal
W}_1}{\cal W}_0=\Tan {\cal W}_1\oplus{\rm V}_{{\cal
W}_1}(\rho_1^0)$, so the additional information comes therefore
from the $\rho_1^0$-vertical vectors, and it is just the holonomic
condition. In fact:

\begin{teor}
Let $\psi_0\colon M\to{\cal W}_0$ be a section fulfilling equation
(\ref{psi0}), $\psi_0=(\psi_\Lag,\psi_{\cal
H})=(\psi_\Lag,\widetilde{\cal FL}\circ\psi_\Lag)$, where
$\psi_\Lag=\rho_1^0\circ\psi_0$. Then:
 \ben
  \item $\psi_\Lag$ is
the canonical lift of the projected section
$\phi=\rho_E^0\circ\psi_0\colon M \to E$ (that is, $\psi_\Lag$ is
a holonomic section).
 \item The section $\psi_\Lag=j^1\phi$ is a
solution to the Lagrangian problem, and the section
$\mu\circ\psi_{\cal H}=\mu\circ\widetilde{\cal
FL}\circ\psi_\Lag={\cal FL}\circ j^1\phi$ is a solution to the
Hamiltonian problem.

Conversely, for every section $\phi\colon M\to E$ such that
$j^1\phi$ is solutions to the Lagrangian problem (and hence ${\cal
FL}\circ j^1\phi$ is solution to the Hamiltonian problem) we have
that the section $\psi_0=(j^1\phi,\widetilde{\cal FL}\circ
j^1\phi)$, is a solution to (\ref{psi0}).
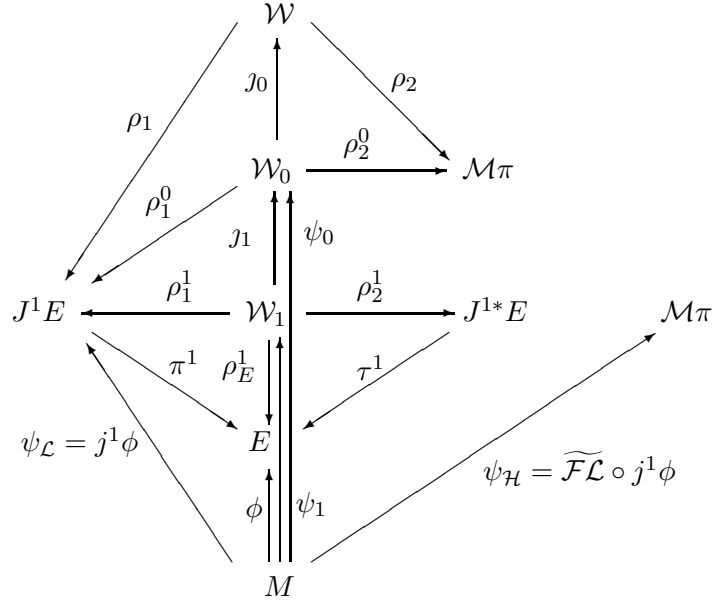
$$
\begin{array}{cccc}
& \begin{picture}(135,100)(0,0)
\put(65,89){\mbox{${\cal W}$}}
\put(13,50){\mbox{$\rho_1$}}
\put(55,89){\vector(-2,-3){65}}
\put(58,65){\mbox{$\jmath_0$}}
\put(70,45){\vector(0,1){38}}
\put(83,89){\vector(1,-1){52}}
\put(113,65){\mbox{$\rho_2$}}
\put(60,30){\mbox{${\cal W}_0$}}
\put(140,30){\mbox{${\cal M}\pi$}}
\put(20,18){\mbox{$\rho^0_1$}}
\put(55,27){\vector(-3,-2){55}}
\put(52,6){\mbox{$\jmath_1$}}
\put(69,-10){\vector(0,1){35}}
\put(81,33){\vector(1,0){53}}
\put(95,40){\mbox{$\rho^0_2$}}
\end{picture} & &
\\
J^1E &
\begin{picture}(135,20)(0,0)
\put(28,10){\mbox{$\rho^1_1$}}
\put(52,3){\vector(-1,0){56}}
\put(58,0){\mbox{${\cal W}_1$}}
\put(100,10){\mbox{$\rho^1_2$}}
\put(81,3){\vector(1,0){56}}
\end{picture}
&
\begin{picture}(80,20)(0,0)
\put(-5,0){\mbox{$J^{1*}E$}}
\put(70,0){\mbox{${\cal M}\pi$}}
\end{picture}
\\ &
\begin{picture}(135,100)(0,0)
\put(29,84){\mbox{$\pi^1$}}
\put(49,84){\mbox{$\rho_E^1$}}
\put(100,82){\mbox{$\tau^1$}}
\put(-27,55){\mbox{$\psi_\Lag=j^1\phi$}}
 \put(148,44){\mbox{$\psi_{\cal H}=\widetilde{\cal FL}\circ j^1\phi$}}
\put(77,32){\mbox{$\psi_1$}}
\put(80,135){\mbox{$\psi_0$}}
 \put(58,30){\mbox{$\phi$}}
\put(59,55){\mbox{$E$}}
 \put(65,0){\mbox{$M$}}
\put(67,97){\vector(0,-1){32}}
\put(0,100){\vector(3,-2){55}}
 \put(135,100){\vector(-3,-2){55}}
\put(53,13){\vector(-2,3){55}}
 \put(83,13){\vector(3,2){130}}
\put(67,13){\vector(0,1){35}}
 \put(71,13){\vector(0,1){85}}
 \put(75,13){\vector(0,1){140}}
\end{picture} &
\begin{picture}(10,100)(0,0)
\end{picture} &
\end{array}
$$
\een
\label{mainteor1}
\end{teor}
\proof
\ben
\item
Taking \dst\left\{\derpar{}{p_A^\alpha}\right\}\)
 as a local basis for the $\rho^0_1$-vertical vector fields:
$$
\inn\left(\derpar{}{p_A^{\alpha}}\right)\Omega_0  =
 v^A_{\alpha}\d^mx-\d y^A\wedge\d^{m-1}x_\alpha
$$
so that for a section $\psi _0$, we have
$$
0=\psi _0^*\left[\inn\left(\derpar{}{p_A^\alpha}\right)\Omega_0\right]=
\left(v^A_\alpha(x)-\derpar{y^A}{x^\alpha}\right)\d^mx
$$
and thus the holonomy condition appears naturally within the
unified formalism,
 and it is not necessary to impose it by hand to $\psi _0$.
Thus we have that
\dst\psi_0=\left(x^\alpha,y^A,\derpar{y^A}{x^\alpha},\derpar{L}{v^A_\alpha}\right)\),
since $\psi_0$ takes values in ${\cal W}_1$, and hence it is of
the form $\psi_0=(j^1\phi,\widetilde{\cal FL}\circ j^1\phi)$, for
$\phi=(x^\alpha,y^A)=\rho_E^0\circ\psi_0$.
 \item
 Since sections
$\psi_0\colon M\to{\cal W}_0$ solution to (\ref{psi0}) take values
in ${\cal W}_1$, we can identify them with sections $\psi_1\colon
M\to{\cal W}_1$. These sections $\psi_1$ verify, in particular,
that $\psi_1^*\inn(Y_1)\Omega_1=0$ holds for every
$Y_1\in\vf({\cal W}_1)$. Obviously $\psi_0=\jmath_1\circ\psi_1$.
Moreover, as ${\cal W}_1$ is the graph of $\widetilde{\cal FL}$,
denoting by $\rho_1^1=\rho_1^0\circ\jmath_1\colon{\cal W}_1 \to
J^1E$ the diffeomorphism which identifies ${\cal W}_1$ with
$J^1E$, if we define $\Omega_1=\jmath_1^*\Omega_0$, we have that
$\Omega_1=\rho_1^{1*}\Omega_\Lag$. In fact; as
$(\rho_1^1)^{-1}(\bar y)=(\bar y,\widetilde{{\cal F}\Lag}(\bar
y))$, for every $\bar y\in J^1E$, then
$(\rho_0^2\circ\jmath_1\circ(\rho_1^1)^{-1})(\bar y)=
\widetilde{{\cal F}\Lag}(\bar y)\in{\cal M}\pi$, and hence
$$
\Omega_\Lag=
(\rho_0^2\circ\jmath_1\circ(\rho_1^1)^{-1})^*\Omega=
[((\rho_1^1)^{-1})^*\circ\jmath_1^*\circ\rho_0^{2*}]\Omega=
[((\rho_1^1)^{-1})^*\circ\jmath_1^*]\Omega_0=
((\rho_1^1)^{-1})^*\Omega_1
$$
Now, let $X\in\vf(J^1E)$. We have
\bea
(j^1\phi)^*\inn(X)\Omega_\Lag&=&
(\rho_1^0\circ\psi_0)^*\inn(X)\Omega_\Lag=
(\rho_1^0\circ\jmath_1\circ\psi_1)^*\inn(X)\Omega_\Lag
\nonumber \\ &=&
(\rho_1^1\circ\psi_1)^*\inn(X)\Omega_\Lag=
\psi_1^*\inn((\rho_1^1)_*^{-1}X)(\rho_1^{1*}\Omega_\Lag)=
\psi_1^*\inn(Y_1)\Omega_1
\nonumber \\ &=&
\psi_1^*\inn(Y_1)(\jmath_1^*\Omega_0)=
(\psi_1^*\circ\jmath_1^*)\inn(Y_0)\Omega_0=
\psi_0^*\inn(Y_0)\Omega_0
\label{chain}
\eea
where $Y_0\in\vf({\cal W}_0)$ is such that $Y_0=\jmath_{1*}Y_1$.
But as $\psi_0^*\inn(Y_0)\Omega_0=0$, for every $Y_0\in\vf({\cal W}_0)$,
then we conclude that $(j^1\phi)^*\inn(X)\Omega_\Lag=0$,
for every $X\in\vf(J^1E)$.

Conversely, let $j^1\phi\colon M\to J^1E$ such that
$(j^1\phi)^*\inn(X)\Omega_\Lag=0$, for every $X\in\vf(J^1E)$,
and define $\psi_0\colon M\to{\cal W}_0$ as
$\psi_0:=(j^1\phi,\widetilde{{\cal F}\Lag}\circ j^1\phi)$
(observe that $\psi_0$ takes its values in ${\cal W}_1$).
Taking into account that, on the points of ${\cal W}_1$,
every $Y_0\in\vf({\cal W}_0)$ splits into
$Y_0=Y_0^1+Y_0^2$, with $Y_0^1\in\vf({\cal W}_0)$ tangent to ${\cal W}_1$, and
$Y_0^2\in\vf^{{\rm V}(\rho_1^0)}({\cal W}_0)$, we have that
$$
\psi_0^*\inn(Y_0)\Omega_0=\psi_0^*\inn(Y_0^1)\Omega_0+\psi_0^*\inn(Y_0^2)\Omega_0=0
$$
because for $Y_0^1$, the same reasoning as in (\ref{chain}) leads to
$$
\psi_0^*\inn(Y_0^1)\Omega_0=(j^1\phi)^*\inn(X_0^1)\Omega_\Lag=0
$$
(where $X_0^1=(\rho_1^1)^{-1}_*Y_0^1$)
and for $Y_0^2$, following also the same reasoning as in (\ref{chain}),
a local calculus gives
$$
\psi_0^*\inn(Y_0^2)\Omega_0=
(j^1\phi)^*\left[\left(f_A^\alpha(x)\left( v_\alpha^A-\derpar{y^A}{x^\alpha}\right)\right)
\d^mx\right]=0
$$
since $j^1\phi$ is a holonomic section.

The result for the sections ${\cal FL}\circ j^1\phi$
is a direct consequence of the {\sl equivalence theorem}
between the Lagrangian and Hamiltonian formalisms
(see, for instance, \cite{EMR-99b} and \cite{EMR-00}).
\qed
\een

{\bf Remark}:
The results in this section can also be recovered in coordinates
taking an arbitrary local vector field
\dst Y_0=f^A\derpar{}{y^A}+g_\alpha^A\derpar{}{v^A_\alpha}+
h^\alpha_A\derpar{}{p_A^\alpha}\in\vf({\cal W}_0)\), then
$$
\inn(Y_0)\Omega_0=
-f^A\derpar{L}{y^A}\d^mx+f^A\d p_A^\alpha\wedge\d^{m-1}x_\alpha +
g_\alpha^A\left(p_A^\alpha-\derpar{L}{v^A_\alpha}\right)\d^mx +
h^\alpha_Av^A_{\alpha}\d^mx-h^\alpha_A\d y^A\wedge\d^{m-1}x_\alpha
$$
and, for a section $\psi_0$ fulfilling (\ref{psi0}),
$$
0=\psi _0^*\inn(Y_0)\Omega_0=
\left[
f^A\left(\derpar{p_A^\alpha}{x^\alpha}-\derpar{L}{y^A}\right)+
g_\alpha^A\left(p_A^\alpha-\derpar{L}{v^A_\alpha}\right)+
h^\alpha_A\left(v^A_\alpha-\derpar{y^A}{x^\alpha}\right)
\right]\d^mx
$$
reproduces the Euler-Lagrange equations,
the restricted Legendre map (that is, the definition of the momenta),
and the holonomy condition.

Summarizing, the equation (\ref{psi0}) gives different kinds of
information, depending on the
 type of verticallity of the vector fields $Y_0$ involved.
In particular we have obtained equations of three different
classes:
 \ben
  \item
 Algebraic (not differential) equations,
determining a subset ${\cal W}_1$ of ${\cal W}_0$, where the
sections solution must take their values. These can be called {\sl
primary Hamiltonian constraints}, and in fact they generate, by
$\hat\rho_2^0$ projection, the primary constraints of the
Hamiltonian formalism for singular Lagrangians, i.e., the image of
the Legendre transformation, ${\cal FL}(J^1E) \subset J^{1*}E$.
\item
 The holonomic differential equations, forcing the sections
solution $\psi_0$ to be lifting of $\pi$-sections. This property
is similar to the one in the unified formalism of Classical
Mechanics, and it reflects the fact that the geometric condition
in the unified formalism is stronger  than the usual one in the
Lagrangian formalism.
 \item
  The classical Euler-Lagrange equations.
 \een

\subsection{The field equations for $m$-vector fields, connections and jet fields}

  The problem of finding sections solution to (\ref{psi0})
  can be formulated equivalently as follows:
  finding a distribution $D_0$ of $\Tan ({\cal W}_0)$ such that
  it is integrable (that is, {\sl involutive\/}),
  $m$-dimensional, $\rho_M^0$-transverse, and
  the integral manifolds of $D_0$ are the sections solution
  to the above equations. (Note that we do not ask them to be lifting
  of $\pi$-sections; that is, the holonomic condition).
  This is equivalent to stating that the sections solution to
  this problem are the integral sections of one of the following
  equivalent elements:
\bit
\item
A class of integrable and $\rho_M^0$-transverse $m$-vector fields
  $\{ X_0\}\subset\vf^m({\cal W}_0)$ satisfying that
  \beq
  \inn (X_0)\Omega_0=0 \quad ,   \quad
  \mbox{\rm for every $X_0\in\{ X_0\}$}
  \label{mvfuf}
  \eeq
\item
An integrable connection $\nabla_0$
in $\rho_M^0\colon{\cal W}_0\to M$ such that
\beq
  \inn(\nabla_0)\Omega_0=(m-1)\Omega_0
\label{ecfuf}
\eeq
\item
An integrable jet field
  ${\mit\Psi}_0\colon{\cal W}_0\to J^1{\cal W}_0$, such that
\beq
  \inn ({\mit\Psi}_0)\Omega_0=0
\label{jfuf}
\eeq
\eit
Locally decomposable and $\rho_M^0$-transverse $m$-vector fields,
 orientable jet fields and orientable connections
  which are solutions of these equations
  will be called {\sl Lagrange-Hamiltonian $m$-vector fields,
  jet fields} and {\sl connections} for $({\cal W}_0,\Omega_0)$.

Recall that, in a natural chart in ${\cal W}_0$,
the local expressions of a connection form, its associated jet field,
and the $m$-multivector fields of the corresponding associated class are
\bea
\nabla_0 &=& \d x^\alpha\otimes
\left(\derpar{}{x^\alpha}+F_\alpha^A\derpar{}{y^A}+
  G_{\alpha\nu}^A\derpar{}{v_\nu^A}+H^\nu_{\alpha A}\derpar{}{p^\nu_A}\right)
\nonumber \\
{\mit\Psi}_0&=&
(x^\alpha,y^A,v^A_\alpha,F^A_\alpha,G^A_{\alpha\eta},H^\nu_{\alpha A})
\nonumber \\
  X_0 &=& f \bigwedge_{\alpha=1}^m
  \left(\derpar{}{x^\alpha}+F_\alpha^A\derpar{}{y^A}+
  G_{\alpha\nu}^A\derpar{}{v_\nu^A}+H^\nu_{\alpha A}\derpar{}{p^\nu_A}\right)
\label{mvf0}
  \eea
  where $f\in\Cinfty(J^1E)$ is an arbitrary non-vanishing function.
  A representative of the class $\{ X\}$ can be selected by the condition
  $\inn (X)(\bar\rho_M^{0*}\omega)=1$, which leads to
  $f=1$ in the above local expression.

Now, the equivalence of the unified formalism with the Lagrangian and Hamiltonian
formalisms can be recovered as follows:

\begin{teor}
Let $\{ X_0\}$ be a class of integrable Lagrange-Hamiltonian
$m$-vector fields in  ${\cal W}_0$, whose elements $X_0\colon{\cal
W}_0\to\Lambda^m\Tan {\cal W}_0$ are solutions of (\ref{mvfuf}),
and let $\nabla_0\colon{\cal W}_0\to
\rho_M^{0*}\Tan^*M\otimes_{{\cal W}_0}\Tan {\cal W}_0$ be its
associated Lagrange-Hamiltonian connection form (which is a
solution to (\ref{ecfuf})),
 and ${\mit\Psi}_0\colon{\cal W}_0\to J^1 {\cal W}_1$ its associated
Lagrange-Hamiltonian jet field (which is a solution to
(\ref{jfuf})).
 \ben
\item
For every $X_0\in\{ X_0\}$, the $m$-vector field $X_\Lag\colon J^1E\to\Lambda^m\Tan J^1E$
defined by
$$
X_\Lag\circ\rho_1^0=\Lambda^m\Tan\rho_1^0\circ X_0
$$
is a holonomic Euler-Lagrange $m$-vector field for the Lagrangian system $\ls$
(where $\Lambda^m\Tan\rho_1^0\colon\Lambda^m\Tan{\cal W}_0\to\Lambda^m\Tan J^1E$
is the natural extension of $\Tan\rho_1^0$).

Conversely, every holonomic Euler-Lagrange $m$-vector field for the
Lagrangian system $\ls$ can be recovered in this way from an integrable
Lagrange-Hamiltonian $m$-vector field $X_0\in\vf^m_{{\cal W}_1}({\cal W}_0)$.
\item
The Ehresmann connection form
$\nabla_\Lag\colon J^1E\to\bar\pi^{1*}\Tan^*M\otimes_{J^1E}\Tan J^1E$
defined by
$$
\nabla_\Lag\circ\rho_1^0=\kappa_{{\cal W}_0}\circ\nabla_0
$$
is a holonomic Euler-Lagrange connection form for the Lagrangian system $\ls$.
(where $\kappa_{{\cal W}_0}$
is defined as the map making the following diagram commutative)
$$
\begin{array}{ccc}
\rho_M^{0*}\Tan^*M\otimes_{{\cal W}_0}\Tan{\cal W}_0
&
\begin{picture}(55,10)(0,0)
\put(15,6){\mbox{$\kappa_{{\cal W}_0}$}}
\put(0,1){\vector(1,0){55}}
\end{picture} &
\bar\pi^{1*}\Tan^*M\otimes_{J^1E}\Tan J^1E
\\
\begin{picture}(15,35)(0,0)
\put(8,35){\vector(0,-1){35}}
\end{picture}
 & &
\begin{picture}(15,35)(0,0)
\put(8,35){\vector(0,-1){35}}
\end{picture}
\\
{\cal W}_0 &
\begin{picture}(55,10)(0,0)
\put(22,6){\mbox{$\rho_1^0$}}
\put(0,1){\vector(1,0){55}}
\end{picture}
& J^1E
\end{array}
$$

Conversely, every holonomic Euler-Lagrange connection form for the
Lagrangian system $\ls$ can be recovered in this way from an integrable
Lagrange-Hamiltonian connection form $\nabla_0$.
\item
The jet field ${\mit\Psi}_\Lag\colon J^1E\to J^1J^1E$ defined by
$$
{\mit\Psi}_\Lag\circ\rho_1^0=j^1\rho_1^0\circ{\mit\Psi}_0
$$
is a holonomic Euler-Lagrange jet field for the Lagrangian system $\ls$.

Conversely, every holonomic Euler-Lagrange jet field for the
Lagrangian system $\ls$ can be recovered in this way from an integrable
Lagrange-Hamiltonian jet field ${\mit\Psi}_0$.
\een
\end{teor}
\proof
 Let $X_0$ be a $\rho_M^0$-transversal $m$-vector field on
${\cal W}_0$ solution to (\ref{mvfuf}). As sections $\psi_0\colon
M \to {\cal W}_0$ solution to the geometric equation (\ref{psi0})
must take value in ${\cal W}_1$, then $X_0$ can be identified with
a $m$-vector field $X_1\colon{\cal W}_0\to\Lambda^m\Tan{\cal W}_1$
(i.e., $\Lambda^m\Tan\jmath_1\circ X_1=X_0\vert_{{\cal W}_1}$),
and hence there exists $X_{\cal L}\colon J^1E\to\Lambda^m\Tan
J^1E$ such that $X_1=\Lambda^m\Tan(\rho_1^1)^{-1}\circ
X_\Lag\in\vf^m({\cal W}_1)$. Therefore, as a consequence of item 1
in theorem \ref{mainteor1},
 for every section $\psi_0$ solution to (\ref{psi0}), there exists
$X_{\cal L}^0\in\vf^m(j^1\phi(M))$ such that
$\Lambda^m\Tan\jmath_{\phi}\circ X_\Lag^0=X_\Lag\vert_{j^1\phi(M)}$,
where $\jmath_\phi\colon j^1\phi\to E$ is the natural imbedding.
So, $X_\Lag$ is $\bar\pi^1$-transversal and holonomic.
 Then, bearing in mind that
$\jmath_1^*\Omega_0=\rho_1^{1*}\Omega_\Lag$, we have
$$
\jmath_1^*\inn(X_0)\Omega_0=\inn(X_1)(\jmath_1^*\Omega_0)=
\inn(X_1)(\rho_1^{1*}\Omega_\Lag)=
\rho_1^{1*}\inn(X_\Lag)\Omega_\Lag
$$
then $\inn(X_0)\Omega_0=0\,\Rightarrow\,\inn(X_\Lag)\Omega_\Lag=0$.

Conversely, given an holonomic Euler-Lagrange $m$-vector field $X_\Lag$,
from $\inn(X_\Lag)\Omega_{\Lag}=0$, and taking into account the
above chain of equalities, we obtain that
$\inn(X_0)\Omega_0\in [\vf({\cal W}_1)]^0$
(the annihilator of $\vf({\cal W}_1)$).
 Moreover, being $X_\Lag$ holonomic, $X_0$ is holonomic,
 and then the extra condition
$\inn(Y_0)\inn(X_0)\Omega_0 = 0$ is also fulfilled for every
$Y_0\in\vf^{{\rm V}(\rho^0_{1*})}({\cal W}_0)$.
 Thus, remembering that
$\Tan_{{\cal W}_1}{\cal W}_0=\Tan {\cal W}_1\oplus{\rm V}_{{\cal W}_1}(\rho_1^0)$,
we conclude that $\inn(X_0)\Omega_0 = 0$.

The proof for Ehresmann connections and jet fields is
straightforward, taking into account that they are equivalent
alternative descriptions in the Lagrangian formalism.
 \qed

This statement also holds for non-integrable classes of $m$-vector
fields, connections and jet fields in ${\cal W}_0$, but now the
corresponding classes of Euler-Lagrange $m$-vector fields,
connections and jet fields in $J^1E$  will not be holonomic (but
only semi-holonomic). To prove this assertion it suffices to
compute the equation (\ref{mvfuf}) in coordinates, using the local
expressions (\ref{coor0}) and (\ref{mvf0}), concluding then that,
in the expressions (\ref{mvf0}), $F_\alpha^A=v_\alpha^A$, which is
the local expression of the semi-holonomy condition (see also
\cite{LMM-2002}).

Finally the Hamiltonian formalism is recovered
in the usual way, by using the following:

\begin{teor}
 Let $\hs$ be the Hamiltonian system associated with
 a (hyper) regular Lagrangian system $\ls$.
 \ben
 \item
{\rm (Equivalence theorem for $m$-vector fields)} Let
$X_{\Lag}\in\vf^m(J^1E)$ and $X_{\cal H}\in\vf^m(J^{1*}E)$ be the
$m$-vector fields solution to the Lagrangian and the Hamiltonian
problems respectively. Then
$$
 \Lambda^m\Tan {\cal F}\Lag\circ X_{\Lag}=fX_{\cal H}\circ {\cal F}\Lag
$$
 for some $f\in\Cinfty (J^{1*}E)$ (we say that the classes $\{ X_{\Lag}\}$ and
 $\{ X_{\cal H}\}$ are ${\cal F}\Lag$-related).
\item
 {\rm (Equivalence theorem for jet fields and connections)} Let
 ${\cal Y}_{\Lag}$ and ${\cal Y}_{\cal H}$ be the jet fields
 solution of the Lagrangian and the Hamiltonian problems respectively. Then
 $$
 j^1{\cal F}\Lag\circ{\cal Y}_{\Lag}={\cal Y}_{\cal H}\circ {\cal F}\Lag
 $$
 (we say that the jet fields ${\cal Y}_{\Lag}$ and
 ${\cal Y}_{\cal H}$ are ${\cal F}\Lag$-related). As a consequence, their
associated connection forms, $\nabla_{\Lag}$ and $\nabla_{\cal H}$
respectively, are ${\cal F}\Lag$-related too.
 \een
(For almost-regular systems the statement is the same, but
changing $J^{1*}E$ for ${\cal P}$).
\end{teor}
\proof
 See \cite{EMR-99b}. (The proof for the almost-regular case
follows in a straightforward way). \qed

As a consequence of these latter theorems, similar comments to
those made at the end of Sections \ref{lagform} and \ref{hamform}
about the existence, integrability and non-uniqueness of
Euler-Lagrange and Hamilton-de Donder-Weyl $m$-vector fields,
connections and jet fields, can be applied to their associated
elements in the unified formalism. In particular, for singular
systems, the existence of these solutions is not assured, except
perhaps on some submanifold ${\cal S}\hookrightarrow{\cal W}_1$,
and the number of arbitrary functions which appear depends on the
dimension of ${\cal S}$ and the rank of the Hessian matrix of $L$
(an algorithm for finding this submanifold is outlined in
\cite{LMM-2002}). The integrability of these solutions is not
assured (even in the regular case), except perhaps on a smaller
submanifold ${\cal I}\hookrightarrow{\cal S}$ such that the
integral sections are contained in ${\cal I}$.

\section{Example: minimal surfaces (in $\Real^3$)}

(In \cite{LMM-2002} we find another interesting example, the {\sl
bosonic string} (which is a singular model), described in this
unified formalism).

\subsection{Statement of the problem. Geometric elements}

The problem consists in looking for mappings
$\varphi\colon U\subset\Real^2\to\Real$ such that their graphs have minimal area
as sets of $\Real^3$, and satisfy certain boundary conditions.

For this model, we have that $M=\Real^2$, $E=\Real^2\times\Real$, and
\beann
J^1E&=&\pi^*\Tan^*\Real^2\otimes\Real=\pi^*\Tan^*M=\pi^*\Tan^*\Real^2
\\
{\cal M}\pi&=&\pi^*(\Tan M\times_ME)
\quad \mbox{\rm (affine maps from $J^1E$ to $\pi^*\Lambda^2\Tan^*M$)}
\\
J^{1*}E&=&\pi^*\Tan M=\pi^*\Tan\Real^2
\quad \mbox{\rm (classes of affine maps from $J^1E$ to $\pi^*\Lambda^2\Tan^*M$)}
\eeann
The coordinates in $J^1E$, $J^{1*}E$ and ${\cal M}\pi$ are denoted
$(x^1,x^2,y,v_1,v_2)$, $(x^1,x^2,y,p^1,p^2)$, and $(x^1,x^2,y,p^1,p^2,p)$
respectively. If $\omega=\d x^1\wedge\d x^2$, the Lagrangian density is
$$
\Lag=[1+(v_1)^2+(v_2)^2]^{1/2}\d x^1\wedge\d x^2\equiv L\d x^1\wedge\d x^2
$$
and the Poincar\'e-Cartan forms are
\beann
\Theta_\Lag&=& \frac{v_1}{L}\d y\wedge\d x^2-\frac{v_2}{L}\d y\wedge\d x^1+
L\left( 1-\left(\frac{v_1}{L}\right)^2-\left(\frac{v_2}{L}\right)^2\right)
\d x^1\wedge\d x^2
\\
\Omega_\Lag&=& -\d\left(\frac{v_1}{L}\right)\wedge\d y\wedge\d x^2+
\d\left(\frac{v_2}{L}\right)\wedge\d y\wedge\d x^1-
\d\left[ L\left( 1-\left(\frac{v_1}{L}\right)^2-\left(\frac{v_2}{L}\right)^2\right)\right]
\wedge\d x^1\wedge\d x^2
\eeann
The Legendre maps are
\beann
{\cal FL}(x^1,x^2,y,v_1,v_2)&=&(x^1,x^2,y,\frac{v_1}{L},\frac{v_2}{L})
\\
\widetilde{\cal FL}(x^1,x^2,y,v_1,v_2)&=&
\left( x^1,x^2,y,\frac{v_1}{L},\frac{v_2}{L},L-\frac{(v_1) ^2}{L}-\frac{(v_2)^2}{L}\right)
\eeann
and then $\Lag$ is hyperregular. The Hamiltonian function is
\beq
H=-[1-(p^1)^2-(p^2)^2]^{1/2}
\label{hamex}
\eeq
So the Hamilton-Cartan forms are
\beann
\Theta_h&=& p^1\d y\wedge\d x^2-p^2\d y\wedge\d x^1-H\d x^1\wedge\d x^2
\\
\Omega_h&=& -\d p^1\wedge\d y\wedge\d x^2+\d p^2\wedge\d y\wedge\d x^1+
\d H\wedge\d x^1\wedge\d x^2
\eeann

\subsection{Unified formalism}

For the unified formalism we have
$$
{\cal W}=\pi^*\Tan^*M\times_E\pi^*(\Tan M\times_ME)
\quad ,\quad
{\cal W}_r=\pi^*\Tan^*M\times_E\pi^*\Tan M=\pi^*(\Tan^*M\times_M\Tan M)
$$
If $w=(x^1,x^2,y,v_1,v_2,p^1,p^2,p)\in{\cal W}$, the coupling form is
$$
\hat{\cal C}=(p^1v_1+p^2v_2+p)\d x^1\wedge\d x^2
$$
therefore
$$
{\cal W}_0=\{ (x^1,x^2,y,v_1,v_2,p^1,p^2,p)\in{\cal W}\ \mid\
[1+(v_1)^2+(v_2)^2]^{1/2}-p^1v_1-p^2v_2-p=0\}
$$
and we have the forms
\beann
\Theta_0&=&
([1+(v_1)^2+(v_2)^2]^{1/2}-p^1v_1-p^2v_2)\d x^1\wedge\d x^2-
p^2\d y\wedge\d x_1+p^1\d y\wedge\d x_2
\\
\Omega_0&=&
-\d ([1+(v_1)^2+(v_2)^2]^{1/2}-p^1v_1-p^2v_2)\wedge\d x^1\wedge\d x^2+
\d p^2\wedge\d y\wedge\d x_1-\d p^1\wedge\d y\wedge\d x_2
\eeann
Taking first $\hat\rho_2^0$-vertical vector fields \dst\derpar{}{v_\alpha}\)
we obtain
$$
0=\inn\left(\derpar{}{v_\alpha}\right)\Omega_0=
\left(p^\alpha-\frac{v_\alpha}{L}\right)\d x^1\wedge\d x^2
$$
which determines the submanifold
${\cal W}_1={\rm graph}\,\widetilde{{\cal FL}}$ (diffeomorphic to $J^1E$),
and reproduces the expression of the Legendre map.
Now, taking $\rho^0_1$-vertical vector fields \dst\derpar{}{p^\alpha}\),
the contraction \dst\inn\left(\derpar{}{p^\alpha}\right)\Omega_0\) gives,
for $\alpha=1,2$ respectively,
$$
 v_1\d x^1\wedge\d x^2-\d y\wedge\d x^2 \quad , \quad
 v_2\d x^1\wedge\d x^2+\d y\wedge\d x^1
$$
so that, for a section
$\psi_0=(x^1,x^2,y(x^1,x^2),v_1(x^1,x^2),v_2(x^1,x^2),p^1(x^1,x^2),p^2(x^1,x^2))$
taking values in ${\cal W}_1$, we have that
the condition
\dst\psi _0^*\left[\inn\left(\derpar{}{p^\alpha}\right)\Omega_0\right]=0\)
leads to
$$
\left(v_1-\derpar{y}{x^1}\right)\d x^1\wedge\d x^2=0 \quad ,\quad
\left(v_2-\derpar{y}{x^2}\right)\d x^1\wedge\d x^2=0
$$
which is the holonomy condition. Finally, taking the vector field \dst\derpar{}{y}\) we have
$$
\inn\left(\derpar{}{y}\right)\Omega_0=-\d p^2\wedge\d x^1+\d p^1\wedge\d x^2
$$
and, for a section $\psi_0$ fulfilling the former conditions,
the equation \dst 0=\psi_0^*\left[\inn\left(\derpar{}{y}\right)\Omega_0\right]\) leads to
\beann
0&=&\left(\derpar{p^2}{x^2}+\derpar{p^1}{x^1}\right)\d x^1\wedge\d x^2=
\left[\derpar{}{x^1}\left(\frac{v_1}{L}\right)+\derpar{}{x^2}\left(\frac{v_2}{L}\right)\right]
\d x^1\wedge\d x^2
\\ &=&
\frac{1}{L^3}
\left[\left( 1+\left(\derpar{y}{x^1}\right)^2\right)\frac{\partial^2y}{\partial x^2\partial x^2}
+\left( 1+\left(\derpar{y}{x^2}\right)^2\right)\frac{\partial^2y}{\partial x^1\partial x^1}-
2\derpar{y}{x^1}\derpar{y}{x^2}\frac{\partial^2y}{\partial x^1\partial x^2}\right]
\d x^1\wedge\d x^2
\eeann
which gives the Euler-Lagrange equation of the problem.

Now, bearing in mind (\ref{hamex}), and the expression of the Legendre map,
from the Euler-Lagrange equations we get
$$
\derpar{y}{x^1}=-\frac{p^1}{H}\quad , \quad
\derpar{y}{x^2}=-\frac{p^2}{H}\quad ; \quad
\derpar{p^1}{x^1}=-\derpar{p^2}{x^2}
$$
which are the Hamilton-De Donder-Weyl equations of the problem.

The $m$-vector fields, connections and jet fields which are the
solutions to the problem in the unified formalism are
 \beann
   X_0 &=& f
  \left(\derpar{}{x^1}+v_1\derpar{}{y}+
  \derpar{v_1}{x^1}\derpar{}{v_1}+\derpar{v_2}{x^1}\derpar{}{v_2}+
  \derpar{p^1}{x^1}\derpar{}{p^1}+\derpar{p^2}{x^1}\derpar{}{p^2}\right)\wedge
\\ & &
  \left(\derpar{}{x^2}+v_2\derpar{}{y}+
  \derpar{v_1}{x^2}\derpar{}{v_1}+\derpar{v_2}{x^2}\derpar{}{v_2}+
  \derpar{p^1}{x^2}\derpar{}{p^1}+\derpar{p^2}{x^2}\derpar{}{p^2}\right)
   \\
{\mit\Psi}_0 &=&
\left( x^1,x^2,y,p^1,p^2;v_1,v_2,
\derpar{v_1}{x^1},\derpar{v_1}{x^2},\derpar{v_2}{x^1},\derpar{v_2}{x^2},
\derpar{p^1}{x^1},\derpar{p^1}{x^2},\derpar{p^2}{x^1},\derpar{p^2}{x^2}\right)
  \\
\nabla_0 &=& \d x^1\otimes
\left(\derpar{}{x^1}+v_1\derpar{}{y}+
\derpar{v_1}{x^1}\derpar{}{v_1}+\derpar{v_2}{x^1}\derpar{}{v_2}+
\derpar{p^1}{x^1}\derpar{}{p^1}+\derpar{p^2}{x^1}\derpar{}{p^2}\right)+
\\ & &
\d x^2\otimes\left(\derpar{}{x^2}+v_2\derpar{}{y}+
\derpar{v_1}{x^2}\derpar{}{v_1}+\derpar{v_2}{x^2}\derpar{}{v_2}+
\derpar{p^1}{x^2}\derpar{}{p^1}+\derpar{p^2}{x^2}\derpar{}{p^2}\right)
\eeann
($f$ being a non-vanishing function)
where the coefficients \dst\derpar{v_\alpha}{x^\nu}=
\frac{\partial^2 y}{\partial x^\nu\partial x^\alpha}\)
are related by the Euler-Lagrange equations, and the coefficients
\dst\derpar{p^\alpha}{x^\nu}\) are related by the Hamilton-De Donder-Weyl equations
(the third one).
Hence the associated Euler-Lagrange $m$-vector fields, connections and jet fields
 which are the solutions to the Lagrangian problem are
\beann
   X_\Lag &=& f
  \left(\derpar{}{x^1}+v_1\derpar{}{y}+
  \derpar{v_1}{x^1}\derpar{}{v_1}+\derpar{v_2}{x^1}\derpar{}{v_2}\right)\wedge
  \left(\derpar{}{x^2}+v_2\derpar{}{y}+
  \derpar{v_1}{x^2}\derpar{}{v_1}+\derpar{v_2}{x^2}\derpar{}{v_2}\right)
   \\
{\mit\Psi}_\Lag &=&
\left( x^1,x^2,y,p^1,p^2;v_1,v_2,
\derpar{v_1}{x^1},\derpar{v_1}{x^2},\derpar{v_2}{x^1},\derpar{v_2}{x^2}\right)
  \\
\nabla_\Lag &=& \d x^1\otimes
\left(\derpar{}{x^1}+v_1\derpar{}{y}+
\derpar{v_1}{x^1}\derpar{}{v_1}+\derpar{v_2}{x^1}\derpar{}{v_2}\right)+
\\ & &
\d x^2\otimes
\left(\derpar{}{x^2}+v_2\derpar{}{y}+
\derpar{v_1}{x^2}\derpar{}{v_1}+\derpar{v_2}{x^2}\derpar{}{v_2}\right)
\eeann
and the corresponding Hamilton-De Donder-Weyl $m$-vector fields, connections and jet fields
 which are the solutions to the Hamiltonian problem are
\beann
   X_{{\cal H}} &=& f
  \left(\derpar{}{x^1}-\frac{p^1}{H}\derpar{}{y}+
  \derpar{p^1}{x^1}\derpar{}{p^1}+
  \derpar{p^2}{x^1}\derpar{}{p^2}\right)\wedge
  \left(\derpar{}{x^2}-\frac{p^2}{H}\derpar{}{y}+
  \derpar{p^1}{x^2}\derpar{}{p^1}+
  \derpar{p^2}{x^2}\derpar{}{p^2}\right)
   \\
{\mit\Psi}_{{\cal H}} &=&
\left( x^1,x^2,y,p^1,p^2;-\frac{p^1}{H},-\frac{p^2}{H},
\derpar{p^1}{x^1},\derpar{p^1}{x^2},\derpar{p^2}{x^1},\derpar{p^2}{x^2}\right)
  \\
\nabla_{{\cal H}} &=& \d x^1\otimes
\left(\derpar{}{x^1}-\frac{p^1}{H}\derpar{}{y}+
\derpar{p^1}{x^1}\derpar{}{p^1}+
\derpar{p^2}{x^1}\derpar{}{p^2}\right)+
\\ & &
\d x^2\otimes
\left(\derpar{}{x^2}-\frac{p^2}{H}\derpar{}{y}+
  \derpar{p^1}{x^2}\derpar{}{p^1}+
  \derpar{p^2}{x^2}\derpar{}{p^2}\right)
\eeann

\section{Conclusions and outlook}

We have generalized the {\sl Rusk-Skinner unified formalism} to
first-order classical field theories. Corresponding to the Whitney
sum $\Tan Q\times_Q\Tan^*Q$ in autonomous mechanics, here we take
 $J^1E\times_E{\cal M}\pi$ as standpoint, but the field equations
are stated in a submanifold ${\cal W}_0\subset J^1E\times_E{\cal
M}\pi$. As a particular case of this situation, the unified
formalism for non-autonomous mechanics is recovered, the Whitney
sum being now $J^1E\times_E\Tan^*E$, where $\pi\colon E\to\Real$
is the configuration bundle \cite{CMC-2002}, \cite{LMM-2002}. Once
the suitable (pre) multisymplectic structures are introduced, the
field equations can be written in several equivalent ways: using
sections and vector fields (\ref{psi0}) in ${\cal W}_0$,
$m$-vector fields (\ref{mvfuf}), connections (\ref{ecfuf}) or jet
fields (\ref{jfuf}).

Starting from equation (\ref{psi0}), we have seen how, when
different kinds of vertical vector fields in ${\cal W}_0$ are
considered, this equation gives a different type of information.
In particular, using $\hat\rho_2^0$-vertical vector fields, we can
define a submanifold ${\cal W}_1\hookrightarrow{\cal W}_0$, which
turns out to be the graph of the (extended) Legendre
transformation (and hence diffeomorphic to $J^1E$). Furthermore,
the field equations are only compatible in ${\cal W}_1$. As
sections solution to the field equations take values in ${\cal
W}_1$, they split in a natural way into two components,
$\psi_0=(\psi_\Lag,\psi_{\cal H})$, (with $\psi_\Lag\colon M \to
J^1E$, and $\psi_{\cal H}=\widetilde{\cal FL}\circ\psi_\Lag$).
Then, taking $\rho_1^0$-vertical vector fields in (\ref{psi0}), we
have proved that the sections solution to the field equations in
the unified formalism are automatically holonomic, even in the
singular case. They are so in the following sense: for every
section $\psi_0$ solution in the unified formalism, the
corresponding section $\psi_\Lag$ is holonomic. (As a special
case, non integrable $m$-vector fields, connections and jet fields
which are solutions to the field equations are semi-holonomic).
These solutions only exist in general in a submanifold of ${\cal
W}_1$. Finally, considering (\ref{psi0}) for a generic vector
field, the Euler-Lagrange equations for $\psi_\Lag$, and the
Hamilton-De Donder-Weyl equations for $\mu\circ\psi_{\cal H}={\cal
F}\Lag\circ\psi_\Lag$
 arise in a natural way. Conversely,
starting from sections $\psi_\Lag=j^1\phi$ and ${\cal
F}\Lag\circ\psi_\Lag$ solutions to the corresponding field
equations, we can recover sections $\psi_0$ solution to
(\ref{psi0}). Thus, we have shown the equivalence between the
standard Lagrangian and Hamiltonian formalisms and the unified
one. This equivalence has been also proved for $m$-vector fields,
connections and jet fields.

Although the subject is not considered in this work, $\cal K$
operators (i.e., the analogous operators in field theories to the
so-called {\sl evolution operator} in mechanics), in their
different alternative definitions \cite{EMMR-01}, can easily be
recovered from the unified formalism, similarly to the case of
classical mechanics.

In a forthcoming paper, this formalism will be applied to give a
geometric framework for Optimal Control with partial differential
equations. Although this subject has been dealt with in the
context of functional analysis, to our knowledge there has been no
 geometric treatment of it to date.

\subsection*{Acknowledgments}

We acknowledge the financial support of {\sl Ministerio de Ciencia y Tecnolog\'\i a},
BFM2002-03493 and BFM2000-1066-C03-01.
We thank Mr. Jeff Palmer for his
assistance in preparing the English version of the manuscript.

\section*{Appendix: $m$-vector fields, jet fields and connections in jet bundles}
\protect\label{mvfdm}

(See \cite{EMR-98} and \cite{LMM-96} for the proofs and other details
  of the following assertions).

Let $E$ be a $n$-dimensional differentiable manifold. For $m < n$,
sections of $\Lambda^m(\Tan E)$ are called $m$-{\sl vector fields}
in $E$ (they are contravariant skew-symmetric tensors of order $m$
in $E$).
  We denote by $\vf^m (E)$ the set of $m$-vector fields in $E$.
  $Y\in\vf^m(E)$ is said to be {\sl locally decomposable} if,
for every $p\in E$, there exists an open neighbourhood $U_p\subset E$
and $Y_1,\ldots ,Y_m\in\vf (U_p)$ such that
$Y\feble{U_p}Y_1\wedge\ldots\wedge Y_m$.
Contraction of $m$-vector fields and tensor fields in $E$ is the usual one.

We can define the following equivalence relation: if
$Y,Y'\in\vf^m(E)$ are non-vanishing $m$-vector fields, then $Y\sim
Y'$ if there exists a non-vanishing function $f\in\Cinfty (E)$
such that $Y'=fY$ (perhaps only in a connected open set
$U\subseteq E$). Equivalence classes will be denoted by $\{ Y\}$.
There is a one-to-one correspondence between the set of
$m$-dimensional orientable distributions $D$ in $\Tan E$ and the
set of the equivalence classes $\{ Y\}$ of non-vanishing, locally
decomposable $m$-vector fields in $E$. Then, there is a bijective
correspondence between the set of classes of locally decomposable
and $\pi$-transverse $m$-vector fields $\{ Y\}\subset\vf^m(E)$,
and the set of orientable jet fields ${\mit\Psi}\colon E\to J^1E$;
that is, the set of orientable Ehresmann connection forms $\nabla$
in $\pi\colon E\to M$. This correspondence is characterized by the
fact that the horizontal subbundle associated with ${\mit\Psi}$
(and $\nabla$) coincides with ${\cal D}(Y)$.

If $Y\in\vf^m(E)$ is non-vanishing and locally decomposable, the
distribution associated with the class $\{ Y\}$ is denoted ${\cal
D}(Y)$. A non-vanishing, locally decomposable $m$-vector field
$Y\in\vf^m(E)$ is said to be {\sl integrable} (resp. {\sl
involutive}) if its associated distribution ${\cal D}_U(Y)$ is
integrable (resp. involutive). Of course, if $Y\in\vf^m(E)$ is
integrable (resp. involutive), then so is every $m$-vector field
in its equivalence class $\{ Y\}$, and all of them have the same
integral manifolds. Moreover, {\sl Frobenius' theorem} allows us
to say that a non-vanishing and locally decomposable $m$-vector
field is integrable if, and only if, it is involutive. Of course,
the orientable jet field ${\mit\Psi}$, and the connection form
$\nabla$ associated with $\{ Y\}$ are integrable if, and only if,
so is $Y$, for every $Y\in\{ Y\}$.

Let us consider the following situation: if $\pi\colon E\to M$ is a fiber bundle,
we are concerned with the case where the integral manifolds of
integrable $m$-vector fields in $E$ are sections of $\pi$.
Thus, $Y\in\vf^m(E)$ is said to be {\sl $\pi$-transverse}
if, at every point $y\in E$,
$(\inn (Y)(\pi^*\beta))_y\not= 0$, for every $\beta\in\df^m(M)$
such that $\beta (\pi(y))\not= 0$.
Then, if $Y\in\vf^m(E)$ is integrable, it is $\pi$-transverse iff
its integral manifolds are local sections of $\pi$.
In this case, if $\phi\colon U\subset M\to E$
is a local section with $\phi (x)=y$ and $\phi (U)$ is
the integral manifold of $Y$ through $y$,
then $\Tan_y({\rm Im}\,\phi)$ is ${\cal D}_y(Y)$.
Integral sections $\phi$ of the class $\{ Y\}$ can be characterized by the condition
$\Lambda^m\Tan\phi=fY\circ\phi\circ\sigma_M$,
where $\sigma_M\colon\Lambda^m\Tan M\to M$ is the natural projection,
and $f\in\Cinfty (E)$ is a non-vanishing function.

As a particular case,
let $\{ X\}\colon J^1E\to D^m\Tan J^1E\subset\{\Lambda^m\Tan J^1E\}$
be a class of non-vanishing, locally decomposable
and $\bar\pi^1$-transverse $m$-vector fields in $J^1E$,
${\mit \Psi}\colon J^1E\to J^1J^1E$ its associated jet field,
and $\nabla\colon J^1E\to\bar\pi^{1*}\Tan M\otimes_{J^1E}\Tan J^1E$
its associated connection form.
Then, these elements are said to be {\sl holonomic} if
they are integrable and their integral sections
$\varphi\colon M\to J^1E$ are holonomic.
Furthermore, consider the $(1,m)$-tensor field in $J^1E$ defined by
  ${\cal J}:=\inn({\cal V})(\bar\pi^{1*}\omega)$,
whose local expression is
\dst {\cal J}=(\d y^A-v^A_\alpha\d x^\alpha)\wedge\d^{m-1}x_\nu
\otimes\derpar{}{v^A_\nu}\) .
A connection form $\nabla$ in $\bar\pi^1\colon J^1E\to M$
(and its associated jet field ${\mit\Psi}\colon J^1E\to J^1J^1E$)
are said to be {\sl semi-holonomic}
(or a {\sl Second Order Partial Differential Equation}), if
${\cal J}(\overbrace{{\rm h}^\nabla,\ldots ,{\rm h}^\nabla}^m)=0$,
where ${\rm h}^\nabla$ denotes the horizontal projector associated with
$\nabla$.
If $\{ X\}\subset\vf^m(J^1E)$ is the associated class of
$\bar\pi^1$-transverse multivector fields, then this condition
is equivalent to
${\cal J}(X)=0$, for every $X\in\{ X\}$.
Then the class $\{ X\}$,
and its associated jet field ${\mit\Psi}$ and connection form $\nabla$
are holonomic if, and only if, they are integrable and semi-holonomic.

\end{document}